\documentclass[11pt]{article} 
\usepackage[text={6.5in,9in}]{geometry}
\usepackage{amsmath,latexsym,amssymb,color,amsthm}
\usepackage{ifthen,graphics,epsfig}
\usepackage{color}
\usepackage{xspace}
\usepackage{enumitem}
\usepackage{url}
\usepackage{amssymb}
\usepackage{pifont}
\usepackage{authblk}
\usepackage{booktabs}
\usepackage{wrapfig}
\usepackage{multirow}
\usepackage[colorlinks=true,citecolor=gray]{hyperref}
\usepackage{caption}
\usepackage{subcaption}
\usepackage[table]{xcolor}

\usepackage{float}
\newfloat{algorithm}{thp}{lop}
\floatname{algorithm}{Algorithm}
\newcommand{\Xomit}[1]{}

\usepackage{marginnote}
\usepackage[table]{xcolor}
\usepackage{todonotes}
\usepackage{menukeys}

\newcommand{\remove}[1]{}

\newlength {\squarewidth}

\newcommand{\toto}{xxx}

\newcounter{linecounter}
\newcommand{\linenumbering}{\ifthenelse{\value{linecounter}<10}{(0\arabic{linecounter})}{(\arabic{linecounter})}}

\renewcommand{\thelinecounter}{\ifnum \value{linecounter} > 9\else 0\fi \arabic{linecounter}}

\newcommand{\numfaults}{\lfloor \frac{n-1}{3} \rfloor}

\newcommand{\prevote}{\textsc{pre-vote}}
\newcommand{\mainvote}{\textsc{main-vote}}
\newcommand{\auxm}{\textsc{auxm}}
\newcommand{\coinecho}{\textsc{coin-echo}}
\newcommand{\auxboth}{\textsc{aux-both}}
\newcommand{\sval}{\textsc{s\_val}}
\newcommand{\svalone}{\textsc{s\_val-s1}}
\newcommand{\svaltwo}{\textsc{s\_val-s2}}
\newcommand{\auxone}{\textsc{aux-stage1}}
\newcommand{\auxstwo}{\textsc{aux-stage2}}

\newcommand{\stage}{\textsc{stage}}

\newcommand{\send}{\mathit{\sf send}}
\newcommand{\sto}{\mathit{\sf to}}
\newcommand{\receive}{\mathit{\sf receive}}
\newcommand{\broadcast}{\mathit{\sf broadcast}}

\newenvironment{smallenum}{
\begin{itemize}[leftmargin=*]
    \setlength{\topsep}{-3pt} 
    \setlength{\partopsep}{-3pt}
  \setlength{\itemsep}{-1pt}
  \setlength{\parskip}{-1pt}
  \setlength{\parsep}{-6pt}
}{\end{itemize}}

\title{Experimental Evaluation of Asynchronous Binary Byzantine Consensus Algorithms with $t < n/3$ and $O(n^2)$ Messages and
  $O(1)$ Round Expected Termination}

\author[1]{Tyler Crain}
\affil[1]{tcrainwork@gmail.com}

\begin{document}

\maketitle

\begin{abstract}
  This work performs an experimental evaluation of four asynchronous binary Byzantine consensus algorithms~\cite{CKS05, C20, C320}
  in various configurations.
  In addition to being asynchronous these algorithms run in rounds, tolerate up to one third of faulty nodes,
  use $O(n^2)$ messages, and use randomized common coins to terminate in an expected constant number of rounds.
  Each of the four algorithms have different requirements for the random coin, for the number of messages needed per round,
  whether or not cryptographic signatures are needed, among other details.
  Two different non-interactive threshold common coin implementations are tested, one using threshold signatures, and one
  based on the Diffe-Hellman problem using validity proofs~\cite{CKS05}.

  Experiments are run in single data center and geo-distributed configurations with between
  $4$ and $48$ nodes. Various simple faulty behaviors are tested.
  As no algorithm performs best in all experimental conditions, two new algorithms introduced that simply combine
  properties of the existing algorithms with the goal of having good performance in the majority of experimental
  settings.

\end{abstract}













\section{Introduction and related work.}
Binary byzantine consensus concerns the problem of getting a set of distinct processes
distributed across a network to agree on a single binary value $0$ or $1$ where processes
can fail in arbitrary ways.
It is well known that this problem is impossible in an asynchronous network with at least
one faulty process~\cite{FLP85}. To get around this, algorithms can employ
randomization, 
or rely on an additional synchrony assumption~\cite{DDS87, DLS88}.
Randomized algorithms largely rely on the existence of a local or common random coin.
The output of local coin is only visible to an individual process, while the output of a common
coin is visible to all processes, but only once a threshold of processes have participated in computing the coin.
A strong common coin is one that outputs the same value at all processes while a weak one
may output different values at different processes with a fixed probability~\cite{R83}.
Let $t$ denote the number of faulty processes in the system.
While algorithms that use synchrony for termination have a lower bound of $(t+1)$ communication steps~\cite{FL82}
for termination, randomized algorithms that use common coins are able to terminate in an expected constant
number of communication steps.

This work presents an experimental evaluation of six different asynchronous randomized binary consensus
algorithms.
It specifically focuses on algorithms that use a predetermined set of participates connected through asynchronous reliable channels
and executes a series of rounds consisting of $1$ or more all to all broadcasts.
Furthermore, the algorithms tolerate up to one third failures (a well know lower bound~\cite{LSP82}), use $O(n^2)$ messages per round,
and have expected termination in a constant number of rounds.
A total of six algorithms are tested, four from previous works and two new algorithms that are introduced here.
Each of the four previous algorithms have different properties that make them attractive in different
conditions. The two new algorithms are simply combinations of the exiting algorithms designed
to perform well in various conditions.

The algorithms can be split into two categories, those that use cryptographic signatures
with the messages sent by the consensus and those that do not.
Generally, the algorithms that use cryptographic signatures have the benefit of requiring fewer message broadcasts per round,
but have the additional network and processing overhead.
Within the algorithms that use cryptographic signatures,~\cite{C220} uses a single broadcast per round and can only decide
the value output by the coin in a round, while~\cite{CKS05} performs $2$ message broadcasts per round and can decide either $0$ or $1$ in each round.
For the algorithms that do note use cryptographic signatures,~\cite{C320} uses $1$-$2$ message broadcasts per round
and can only decide the value output by the coin in a round,
while~\cite{C320} performs $4$ to $5$ message broadcasts per round and can decide either $0$ or $1$ in each round.
The algorithms of~\cite{CKS05, C220, C320}, require a strong common coin for expected constant time termination, while~\cite{C320} may use either
a strong or weak common coin.

While there are many other asynchronous randomized binary consensus algorithms that have been presented in literature~\cite{A03, BO83, B87, BT83, CR93, FP90, KS16, MMR14, PCR14, R83, T84},
the four algorithms used here have been chosen for their efficiency.
The algorithms of~\cite{BO83, B84, KS16} use local coins, but~\cite{BO83, B84} have exponential expected number of rounds,
and~\cite{KS16} has polynomial number of expected rounds.
The algorithms of~\cite{BG93,FMR05,R83} use a common coin, but tolerate $t < n/5$ or fewer faults.
The algorithms of~\cite{B83, ST87, T84} use $O(n^3)$ messages per round.
The algorithm of~\cite{CGR11} uses $O(n^2)$ messages per round,
but has large bit complexity in the size of messages.
The algorithm of~\cite{MMR14} has similar properties to~\cite{C320},
but uses a fair scheduler to ensure termination and may use additional message broadcasts per round.
The algorithm of~\cite{C320} is based on the algorithm of~\cite{MR17},
but uses fewer message broadcasts per round.
The algorithm presented in~\cite{CKS20}, has similar properties to~\cite{C320}, including supporting the use of weak common
coins, but follows a similar design as~\cite{MMR14}, limiting the power of the adversary to reorder messages.

Two different randomized strong common coins are implemented.
The first uses BLS-threshold~\cite{BLS04} signatures and the second~\cite{CKS05} is a threshold coin tossing scheme
using Shamir's secret sharing~\cite{S79} and non-interactive proofs of equality of discrete logarithms~\cite{FS86}.
Both implementations require an initial trusted setup that once performed allows the coin
to be used a polynomial number of times.
While the algorithm of~\cite{C320} supports a weak common coin, we are not aware of any existing weak common coin
schemes that have any advantages over the strong common coins in the asynchronous model.
A weak common coin is presented in~\cite{CKS20} using verifiable random functions~\cite{MRV99}, but the model considered
in that work limits the power of the network adversary to reorder certain messages.

A previous study~\cite{MNCV06} compared the algorithm of~\cite{CKS05} against the local coin algorithm of~\cite{B84}
in a local network. Several other works~\cite{DRZ18,MXC16} have tested the algorithm of~\cite{MMR14}
when used within multi-valued reductions. 

While the binary consensus problem only allows process to agree on a single binary value,
there exist many reductions to multi-value consensus~\cite{MR17, MRT00, TC84, ZC09} allowing processes to agree on arbitrary values.
Furthermore many algorithms~\cite{BSA14, CL02} exists that solve multi-value consensus directly
through the use of types of synchrony assumptions to ensure termination.

\section{A Byzantine Computation Model.}
\label{sec:model}

This section describes the assumed computation model.

\paragraph{Asynchronous processes.}
The system is made up of a set $\Pi$ of $n$ asynchronous sequential processes,
namely $\Pi = \{p_1,\ldots,p_n\}$; $i$ is called the ``index'' of $p_i$. 
``Asynchronous'' means that each process proceeds at its own speed,
which can vary with time and remains unknown to the other processes.
``Sequential'' means that a process executes one step at a time.
This does not prevent it from executing several threads with an appropriate
multiplexing. 
%
Both notations
$i\in Y$ and $p_i\in Y$ are used to say that $p_i$ belongs to the set $Y$.

\paragraph{Communication network.}
\label{sec:basic-comm-operations}
The processes communicate by exchanging messages through
an asynchronous reliable point-to-point network. ``Asynchronous''  means that
there is no bound on message transfer delays, but these delays are finite.
``Reliable'' means that the network does not lose, duplicate, modify, or
create messages. ``Point-to-point'' means that any pair of processes
is connected by a bidirectional channel.
%
A process $p_i$ sends a message to a process $p_j$ by invoking the primitive 
``$\send$ {\sc tag}$(m)$ $\sto~p_j$'', where {\sc tag} is the type
of the message and $m$ its content. To simplify the presentation, it is
assumed that a process can send messages to itself. A process $p_i$ receives 
a message by executing the primitive ``$\receive()$''.
The macro-operation $\broadcast$ {\sc tag}$(m)$ is  used as a shortcut for
``{\bf for each} $p_i \in \Pi$  {\bf do} $\send$ {\sc tag}$(m)$ $\sto~p_j$
{\bf end for}''. 

\paragraph{Failure model.}
Up to $t$ processes can exhibit a {\it Byzantine} behavior~\cite{PSL80}.
 A Byzantine process is a process that behaves
arbitrarily: it can crash, fail to send or receive messages, send
arbitrary messages, start in an arbitrary state, perform arbitrary state
transitions, etc. Moreover, Byzantine processes can collude 
to ``pollute'' the computation (e.g., by sending  messages with the same 
content, while they should send messages with distinct content if 
they were non-faulty). 
A process that exhibits a Byzantine behavior is called {\it faulty}.
Otherwise, it is {\it non-faulty}.  
%
Moreover, it is assumed that the Byzantine processes do not
fully control the network in that they can not corrupt the messages sent by 
non-faulty  processes.
Byzantine processes can control the network by modifying
the order in which messages are received, but they cannot
postpone forever message receptions.  

For this work it is assumed that $t = \numfaults$.

\paragraph{A Threshold Common Coin.}
The model is enriched with the same \emph{threshold common coin} (CC)~\cite{R83}.
The common coin outputs a binary value at each non-faulty process
for each round.
All non-faulty processes output $0$ in round $r$ with probability $1/d$
and output $1$ in round $r$ with probability $1/d$.
Non-faulty processes output different values in round $r$ with probability $(d-2)/d$,
where $d \geq 2$ is a known constant.
The output of the coin is revealed by calling a function {\sf random}() provided by
a random oracle.
The output of the coin is unpredictable and random and its output is only revealed
for a round $r$ once at least some threshold of processes has called {\sf random}() in that round.
This threshold is defined by the requirements of the consensus algorithms.

\paragraph{A Strong Threshold Common Coin.}
A \emph{strong threshold common coin} (SCC) is defined as a threshold common coin where $d = 2$, meaning that in every round
all non-faulty processes receive the same output from the common coin.

\paragraph{Signatures.}
Asymmetric cryptography allow processes to sign messages.
Each process $p_i$ has a public key known by everyone and a private key
known only by $p_i$.
Messages are signed using the private key and can be validated by any process
with the corresponding public key, allowing the process to identify the signer of the message.
Signatures are assumed to be unforgeable.
A process will ignore any message that is malformed or contains an invalid signature.

\paragraph{Non-interactive threshold signatures.}
Given the set of $n$ processes, and a threshold $tr$,
taking $tr$ signatures (called signature shares) of the same message from $tr$ different processes can be combined to
generate a \emph{unique} threshold signature that can be verified by a threshold public key
known by everyone.
Any set of $tr$ signatures of the same message from $tr$ different processes
generates the same threshold signature.
Threshold signatures are assumed to be unforgeable and no set of less than $tr$ nodes
can generate them.





\section{Random threshold common coins.}\label{sec:coin}

This section describes the implementations of the threshold common coins.
The output of a threshold coin is generated for each round of the binary consensus
algorithms.
All implementations rely on a trusted setup~\cite{S79} executed before the consensus
in which a threshold secret key is generated for each of the $n$ consensus participants
for a given threshold $tr$.
For the computation of a coin, a function is defined that takes as input a processes' secret key
and a predefined \emph{coin-message} containing the round and predefined identifier (so that the shares cannot be reused),
as input and outputs a \emph{threshold share} of the coin.
From any set of $tr$ valid threshold shares from a round $r$ the value of the
coin can be computed.

The chosen threshold $tr$ is specified by each consensus algorithm.
Certain algorithms require a threshold of $(t+1)$, while others require
a threshold of $(n-t)$.

While the different coin implementations have similar setup costs and guarantees, the reason for choosing one
implementation over another is primarily based on different computational networking costs.
This is explored in the experiments.
Furthermore, for the coin implementations that use threshold-signatures, the same secret threshold keys are
used to sign the consensus messages of the signature based consensus algorithms.

\subsection{CI:t}
The implementation \emph{CI:t} (coin implementation, threshold-BLS) uses threshold-BLS signatures to generate coin outputs
for the chosen threshold $tr$.
A share of the coin is the threshold signature share of a coin-message for a given round
using a processes' secret key.
Upon reception of $tr$ valid signature shares, the threshold signature is generated
which is then input to a cryptographic hash function, from which the value of the coin
is taken as the first bit output.

\subsection{CI:te}
The implementation \emph{CI:te} (coin implementation, threshold-BLS, echo)
also uses threshold-BLS signatures in the same way as CI:t.
The difference is that CI:te adds an additional message broadcast before non-faulty
processes generate their threshold signature shares for the coin message.
Specifically, each non-faulty process will broadcast a $\coinecho$ message containing the current
round, and after receiving $(n-t)$ of these messages from different processes the process
will participate in generating the coin as in CI:t.
By adding the extra $\coinecho$ message, algorithms that require coin thresholds of
$(n-t)$ can instead using coin thresholds of $(t+1)$.
Given this, CI:te is only used with algorithms that require an $(n-t)$ coin threshold.

The reason for using a $(t+1)$ threshold coin over an $(n-t)$ one is reduced cryptographic computation costs
at the expense of an extra message broadcast.
Note that the $\coinecho$ message does not need to be signed as it is only used to ensure a
that a threshold of non-faulty processes have reached this point in the consensus algorithm
before computing the coin.

\subsection{CI:pc}
The implementation \emph{CI:pc} (coin implementation, proof coin)~\cite{CKS05} is a threshold scheme based
on the Diffie-Hellman problem.
Each non-faulty process generates its share of the coin by raising the hash of a coin-message message for a given round
to the power of its secret key. This is sent along with a proof of validity as a non-interactive proof of equality
of discrete logarithms.
Upon reception of a share, a non-faulty process checks that the validity proof is correct for the share.
Once $tr$ valid shares have been received they are combined using Lagrange interpolation.
The output of this is then input into a cryptographic hash function from which the first bit output is
taken as the value of the coin.
In the random oracle model, this coin-tossing scheme is secure under the
Computational Diffie-Hellman assumption if the threshold is $t+1$,
and under the Decisional Diffie-Hellman assumption otherwise.

\subsection{CI:pce}
The implementation \emph{CI:te} (coin implementation, proof coin, echo) follows the same implementation
as CI:te, except using the Diffie-Hellman based solution instead of the threshold-BLS signatures.
Again the idea here is to allow algorithms that require an $(n-t)$ coin threshold to use
a $(t+1)$ threshold at the cost of an additional $\coinecho$ message.

\section{Binary Byzantine Consensus.}
\label{sec:byz-consensus}
This section describes the Binary Byzantine Consensus problem being solved by the algorithms.

\subsection{The Binary Consensus Problem.}


%

In the binary consensus problem processes input a value to the algorithm, called their \emph{proposal},
run an algorithm consisting of several rounds,
and eventually output a binary value called their \emph{decision}.
Let $\cal V$ be the set of values that can be proposed.
While  $\cal V$ can contain any number ($\geq 2$) of values
in multi-valued consensus, it contains only two values in binary consensus, 
e.g., ${\cal V} =\{0,1\}$.
Assuming that
each non-faulty process proposes a value, the binary Byzantine consensus (BBC) problem is for 
each of them to
decide on a value in such a way that the following properties are
satisfied:
\begin{itemize}
\item BBC-Termination. Every non-faulty process eventually decides on a value.
\item BBC-Agreement.   No two non-faulty processes decide on different values.
\item BBC-Validity.  If all non-faulty processes propose the same value, no
other value can be decided.
\end{itemize}

After a value is decided a non-faulty process may need to continue execution of the consensus
instance to ensure all non-faulty processes decide.
A non-faulty process then \emph{terminates} a consensus instance when it no longer needs
to execute any more operations for the instance given the assumptions of the model.
When considering the implementation, a consensus instance
never really terminates as reliable channels are commonly implemented through message re-transmission,
and the system may need to allow nodes to recover from crash faults through the use of re-transmission.
At a higher level, the application running on top of the consensus may be able to garbage collect
old consensus instances through snapshotting or other coordination mechanisms, but these are not explored
in this work.

\section{Algorithms}\label{sec:alg}
This section briefly describes the construction of the algorithms focusing on their
implementations. The reader is referred to their original papers for details
on the correctness and liveness of the algorithms.

For each algorithm a set of properties are described, the properties are defined as follows:
\fbox{\parbox{\textwidth}{
Description of some key properties used to described the algorithms.
\begin{smallenum}
\item {\bf Consensus messages per round:} The number of consensus messages broadcast in each round in addition to the coin.
\item {\bf Fastest decision:} The number of messages before a non-faulty process can decide in the best case.
\item {\bf Binary values that can be decided in a round:} The possible binary values that can be decided in a singe round.
\item {\bf If the first non-faulty process decides binary value $b$ in round $r$, all non-faulty processes decide at the latest:} After the first non-faulty
  process decides, how long before all non-faulty processes decide.
\item {\bf Weak coin supported:} Whether or not the algorithm can use a weak coin.
\item {\bf Coin threshold:} The number of processes needed to participate in the computation of the coin before the output is revealed.
\item {\bf Signatures:} If the consensus messages need to be signed.
\item {\bf Consensus message contents:} The contents of the consensus messages.
\item {\bf Proofs of validity:} If the consensus messages need to include cryptographic proofs that their contents are valid.
\end{smallenum}
}}

\subsection{BC:s1}
The algorithm \emph{BC:s1}~\cite{C220} (Binary Consensus, signatures, algorithm 1) is a round based
algorithm that requires messages to be signed by cryptographic signatures.
A single message type is used, called an $\auxm$ message containing
a round number and a binary value.

At the start of the algorithm each non-faulty process broadcasts signed $\auxm$ message
with round number $0$ and the processes' initial binary proposal.
After receiving $(n-t)$ of these messages, non-faulty processes execute a series of rounds until termination.

At the start of a round $r$, each non-faulty process $p$ has a \emph{valid} binary value $b$ called its \emph{estimate}.
A binary value $b$ is considered valid in round $r$ if $b$ was proposed by a non-faulty process
and $\neg b$ has not been decided by a non-faulty process in a previous round.
Furthermore the algorithm ensures that when processes $p$ starts round $r$ with estimate $b$ it has
a set of signed $\auxm$ messages from previous rounds proving that $b$ is valid.
Processes $p$ then broadcasts an $\auxm$ message containing the round $r$ and its estimate $b$ plus
the set of signatures proving that $b$ is valid in round $r$.

After receiving $(n-t)$ round $r$ $\auxm$ messages from distinct processes containing valid binary values,
process $p$ then computes the coin for round $r$, which outputs some binary value $c$.
If $p$ had received $(n-t)$ round $r$ $\auxm$ messages with binary value $c$ from distinct processes
then $p$ decides $c$.
Process $p$ then computes its new estimate and continues to round $r+1$.

\paragraph{Proofs of validity.}
As previously mentioned, $\auxm$ messages broadcast by non-faulty processes will
contain a set of signatures proving that the binary value broadcast is valid.
Specifically, the proof will be a set of signatures of a single $\auxm$
message from an earlier round $r'$. If $r' = 0$ then a set of $t+1$ signatures
is needed, otherwise a set of $(n-t)$ signatures is needed. Given that the signatures
will be of the same message and of a specific count, threshold signatures can be used
to reduce the size of the proofs to a single threshold signature.

For the implementation in this paper each process is assigned two threshold keys
during the pre-consensus setup,
one key for a $(t+1)$ threshold and one for a
$(n-t)$ threshold. Round $0$ $\auxm$ messages are signed using the $(t+1)$
threshold key and messages for later rounds are signed with the $(n-t)$ threshold key.

\paragraph{Termination.}
The algorithm ensures that if a non-faulty processes $p$ decides binary value $b$
in round $r$ then all non-faulty processes will, at the latest, decide in the first round
$r' > r$ where the output of the coin in round $r'$ is $b$.
Given that non-faulty processes may decide in different rounds, a processes cannot immediately
terminate as soon as it decides.
While there are different ways ensure all non-faulty processes terminate, the following is implemented:
When a non-faulty process process decides in round $r$, it broadcasts a \emph{proof of decision} containing
the $(n-t)$ round $r$ signed $\auxm$ messages that allowed it to decide, the process then terminates.
This ensures that all other non-faulty processes either decide in round $r'$ or when they receive
this proof (whichever comes first).

\fbox{\parbox{\textwidth}{
    Algorithm BC:s1 properties.
\begin{smallenum}
\item {\bf Consensus message broadcasts per round:} $2$ in the first round, $1$ in following rounds.
\item {\bf Fastest decision:} After $2$ consensus message broadcasts and the computation of a coin.
\item {\bf Binary values that can be decided in a round:} Only the same value as the coin for that round.
\item {\bf If the first non-faulty process decides binary value $b$ in round $r$, all non-faulty processes decide at the latest:} Either of the following
  (whichever happens first): (i) in the first round $r' > r$ where the
  output of the coin in round $r'$ is $b$, or
  (ii) when a proof of decision is received from another process.
\item {\bf Weak coin supported:} No.
\item {\bf Coin threshold:} $(n-t)$.
\item {\bf Signatures:} Required for consensus messages.
\item {\bf Consensus message contents:} Signature, round number, message type identifier, a binary value, and a proof of validity.
\item {\bf Proofs of validity:} Either $(t+1)$ or $(n-t)$ signatures from a previous round, or if using threshold signatures,
  a single threshold signature.
\end{smallenum}
}}

\subsection{BC:s2}
The algorithm \emph{BC:s2}~\cite{CKS05} (Binary Consensus, signatures, algorithm 2) is a round based
algorithm that requires messages to be signed by cryptographic signatures.
The consensus algorithm uses two messages types, a $\prevote$ message containing
a round number and a binary value, and a $\mainvote$ message containing a round
number and either a binary value or $\bot$.

At the start of the algorithm each non-faulty process broadcasts signed $\prevote$ message
with round number $0$ and the processes' initial binary proposal.
After receiving $(n-t)$ of these messages, non-faulty processes execute a series of rounds until termination.

At the start of a round $r$, each non-faulty process $p$ has a \emph{valid} (called \emph{justified} in~\cite{CKS05})
binary value $b$ called its \emph{estimate}.
A binary value $b$ is considered valid in round $r$ if $b$ was proposed by a non-faulty process
and $\neg b$ has not been decided by a non-faulty process in a previous round.
Furthermore the algorithm ensures that when process $p$ starts round $r$ with estimate $b$ it has
a set of signed messages from previous rounds proving that $b$ is valid.
Process $p$ then broadcasts a $\prevote$ message containing the round $r$ and its estimate $b$ plus
the set of signatures proving that $b$ is valid in round $r$.

After receiving $(n-t)$ round $r$ $\prevote$ messages from distinct processes containing valid binary values,
the process then computes the set $bin\_values$ as the set of valid binary values from the $\prevote$ messages.
The process then broadcasts a $\mainvote$ message for round $r$ with either value $\bot$ if $bin\_values = \{0,1\}$,
or with binary value $b$ where $bin\_values = \{b\}$ along with a set of signatures proving
the value is valid.
The value $\bot$ is considered valid in round $r$ if both $0$ and $1$ have been proposed by a non-faulty process and
have not been decided in a previous round.
A binary value $b$ is valid for a $\mainvote$ message in round $r$ if $(n-t)$ $\prevote$ messages are received
from different processes with value $b$.

If a non-faulty process $p$ receives $(n-t)$ round $r$ $\mainvote$ messages with a valid binary value $b$ from distinct processes
then $p$ decides $b$.
Otherwise, after receiving $(n-t)$ round $r$ $\mainvote$ messages from distinct processes containing valid values,
process $p$ computes the coin for round $r$, which outputs some binary value $c$.
The process then uses this to compute its new estimate, and continues to round $r+1$.

\paragraph{Proofs of validity.}
As previously mentioned, $\auxm$ and $\mainvote$ messages broadcast by non-faulty processes will
contain a set of signatures proving that the value broadcast is valid.
For a binary value $b$ the proof will be a set of signatures of a
message with value $b$ from an earlier round $r'$.
If $r' = 0$ then a set of $t+1$ signatures is needed, otherwise a set of $(n-t)$ signatures is needed.
For $\bot$ the proof will be similar, except will contain two sets of signatures, one for $0$ and $1$.
Similar to the BC:s1 algorithm, threshold signatures can be used
to reduce the size of the proofs to a single threshold signature.

As with BC:s1, each process is assigned two threshold keys
during the pre-consensus setup,
one key for the $t+1$ threshold and one for the
$(n-t)$ threshold. Round $0$ $\prevote$ messages are signed using the $(t+1)$
threshold key and messages for later rounds are signed with the $(n-t)$ threshold key.

\paragraph{Termination.}
The algorithm ensures that if a non-faulty processes $p$ decides binary value $b$
in round $r$ then all non-faulty processes will, at the latest, decide in the following round.
Therefore, when a non-faulty process process decides in round $r$, it broadcasts a \emph{proof of decision} containing
the $(n-t)$ round $r$ signed $\mainvote$ messages that allowed it to decide then terminates.
This ensures that all other non-faulty processes either decide in round $r+1$ or when they receive
this proof (whichever comes first).

\fbox{\parbox{\textwidth}{
    Algorithm BC:s2 properties.
\begin{smallenum}
\item {\bf Consensus message broadcasts per round:} $3$ in the first round, $2$ in following rounds.
\item {\bf Fastest decision:} After $3$ consensus message broadcasts.
\item {\bf Binary values that can be decided in a round:} Either $0$ or $1$.
\item {\bf If the first non-faulty process decides binary value $b$ in round $r$, all non-faulty processes decide at the latest:} Either of the following
  (whichever happens first): (i) in round $r+1$, or
  (ii) when a proof of decision is received from another process.
\item {\bf Weak coin supported:} No.
\item {\bf Coin threshold:} $(n-t)$.
\item {\bf Signatures:} Required for consensus messages.
\item {\bf Consensus message contents:} Signature, round number, message type identifier, a constant number of bits, and a proof of validity.
\item {\bf Proofs of validity:} Either $(t+1)$ or $(n-t)$ signatures from a previous round, or if using threshold signatures,
  a single threshold signature. Twice this if the message contains value $\bot$.
\end{smallenum}
}}

\subsection{BC:ns1}
The algorithm \emph{BC:ns1}~\cite{C320} (Binary Consensus, signatures, algorithm 1) is a round based
algorithm that does not require consensus messages to be signed.
It follows a similar design to BC:s1, but may use additional message broadcasts as
a way of replacing the cryptographic proofs of validity used in BC:s1.
The consensus algorithm uses two message types, an $\sval$ message type and an $\auxm$ message type, each containing
a round number and a binary value.

Non-faulty processes execute a series of rounds until termination.
At the start of a round $r$, each non-faulty process $p$ has a valid binary value $b$ called its \emph{estimate}.
A binary value $b$ is considered valid in round $r$ if $b$ was proposed by a non-faulty process
and $\neg b$ has not been decided by a non-faulty process in a previous round.
If the estimate is not equal to the output of the coin from the previous round, then the process
broadcasts an $\sval$ message containing its estimate and the round.
In round $1$ a processes' estimate is its initial binary proposal, which is always
broadcast in an $\sval$ message.

At any point during the algorithm if a non-faulty process receives $t+1$
round $r$ $\sval$ messages from different processes for a binary value $b$ and the process
has not yet broadcast a message containing those same values, the process then broadcasts
such a message.
After receiving $(n-t)$ round $r$ $\sval$ messages with binary value $b$ from different processes,
the value $b$ is considered valid in round $r$.
Broadcasting messages in such a way (referred to as \emph{echoing} a message)
ensures all non-faulty processes eventually agree on the set of valid binary values
in a round.

Once a non-faulty process has computed at least one valid binary value in round $r$, it then broadcasts an
$\auxm$ message for the round containing a valid binary value.
After receiving $(n-t)$ round $r$ $\auxm$ messages from distinct processes containing valid binary values,
the process then computes the coin for round $r$, which outputs some binary value $c$.
If the process had received $(n-t)$ round $r$ $\auxm$ messages with binary value $c$ from distinct processes
then it decides $c$.
A new estimate is computed and the process continues to round $r+1$.

\paragraph{Termination.}
The algorithm ensures that if a non-faulty processes $p$ decides binary value $b$
in round $r$ then all non-faulty processes will, at the latest, decide in the first round
$r' > r$ where the output of the coin in round $r'$ is $b$.
Thus, for the implementation, when a non-faulty process decides in round $r$
it continues executing until round $r'$.

\fbox{\parbox{\textwidth}{
    Algorithm BC:ns1 properties.
\begin{smallenum}
\item {\bf Consensus message broadcasts per round:} $2$ to $3$ in the first round, $1$ to $2$ in following rounds.
\item {\bf Fastest decision:} After $2$ consensus message broadcasts and the computation of a coin.
\item {\bf Binary values that can be decided in a round:} Only the same value as the coin for that round.
\item {\bf If the first non-faulty process decides binary value $b$ in round $r$, all non-faulty processes decide at the latest:} In the first round $r' > r$ where the
  output of the coin in round $r'$ is $b$.
\item {\bf Weak coin supported:} No.
\item {\bf Coin threshold:} $(n-t)$.
\item {\bf Signatures:} Not required for consensus messages.
\item {\bf Consensus message contents:} Round number, message type identifier, and a binary value.
\item {\bf Proofs of validity:} Not required.
\end{smallenum}
}}

\subsection{BC:ns2}
The algorithm \emph{BC:ns1}~\cite{C320} (Binary Consensus, signatures, algorithm 1) is a round based
algorithm that does not require consensus messages to be signed.
While the algorithm is similar to \emph{BC:s2} in that either $0$ or $1$ can be decided
in each round, is uses several additional message broadcasts per round as a way of
replacing the cryptographic proofs of validity used in BC:s2.
It also has the noteworthy properties of supporting strong and weak coins
with a $t+1$ threshold. The process uses five different message types,
$\svalone$, $\auxone$, $\svaltwo$, $\auxstwo$, and $\auxboth$ each containing a round
and a binary value or $\bot$.

As with the other algorithms, non-faulty processes execute a series of rounds until termination.
The design of the algorithm itself is similar to that BC:ns1, where in each round first
an $\sval$ message may be broadcast, followed by an $\auxm$ messages.
Differently than BC:ns1 this pattern is repeated twice per round, with the first set
of $\auxm$ and $\sval$ messages being labeled $\stage0$ and the second set labeled $\stage1$.
Similar to BC:ns1, in order to ensure all non-faulty processes eventually agree on the set of valid binary values
a non-faulty process will echo an $\sval$ message from a stage after receiving the same message from $t+1$
distinct processes.
Additionally, in the middle of the two stages an $\auxboth$ message is broadcast containing the set of binary values that
the process knows is valid in the round.

At the end of the second stage if a non-faulty process receives $(n-t)$ round $r$
$\auxstwo$ messages with valid binary value $b$ from distinct processes then it decides $b$.
Otherwise the process then computes the value of the coin, uses this to compute its new estimate, and continues to the next round.

\paragraph{Termination.}
The algorithm ensures that if a non-faulty processes $p$ decides binary value $b$
in round $r$ then all non-faulty processes will, at the latest, decide in round $r+1$.
Thus, for the implementation, when a non-faulty process decides in round $r$
it continues executing until round $r+1$.

\fbox{\parbox{\textwidth}{
        Algorithm BC:ns2 properties.
\begin{smallenum}
\item {\bf Consensus message broadcasts per round:} $5$ to $6$ in the first round, $4$ to $5$ in following rounds.
\item {\bf Fastest decision:} After $5$ consensus message broadcasts.
\item {\bf Binary values that can be decided in a round:} Either $0$ or $1$
\item {\bf If the first non-faulty process decides binary value $b$ in round $r$, all non-faulty processes decide at the latest:} In round $r+1$.
\item {\bf Weak coin supported:} Yes.
\item {\bf Coin threshold:} $(t+1)$.
\item {\bf Signatures:} Not required for consensus messages.
\item {\bf Consensus message contents:} Round number, message type identifier, and a constant number of bits.
\item {\bf Proofs of validity:} Not required.
\end{smallenum}
}}

\section{Optimizations and additional details.}\label{sec:opt}

\paragraph{Combine coin messages.}
With some care in the way the the estimates of processes are broadcast,
the signature based algorithms can send the coin message of round $r$ and the first consensus message of round
$r+1$ together.
This reduces the number of message steps in rounds following the first round by $1$.
Certain experiments are performed with this optimization.

Note that when the messages are combined, the initial message of round $r+1$ may need to take a special
value (instead of a binary value) saying that the node will support the value of the coin when it is revealed.
In this case a node may not receive $(n-t)$ signature shares of the same message,
and will be unable to generate a threshold signature, thus resulting in needing to send individual signatures
as the proof of validity, increasing network overhead.

\paragraph{No coin after decision.} For the algorithms BC:s2 and BC:ns2, if a non-faulty process decides in round $r$ then
the value of the coin does not need to be computed in round $r$ or any following round. Thus,
once a non-faulty process decides, it does not participate in computing the coin in that or
any following rounds.

\paragraph{Preference towards $1$.}
In the algorithms there are certain cases where both $0$ and $1$ are valid values and a
non-faulty node can choose to broadcast either value. In these cases the implementation
will choose to broadcast $1$. One reason for doing this is that reductions from binary to
multi-value consensus will often use $1$ to represent a value being decided while $0$ will
represent $\bot$ being decided.

\paragraph{Coin presets.} While an unpredictable random threshold coin is needed to ensure termination with probability $1$ in asynchronous networks,
there are many cases where the algorithms can terminate even if the value of the coin is known apriori.
Furthermore if the coin is predefined for certain rounds then the cryptographic computations and message broadcasts
needed to compute the coin in those rounds is not needed.
To test the affect of this in certain experiments the value of the coin will be predefined as having value
$1$ in round $1$ and $0$ in round $2$.
This is particularly effective for the algorithms BC:s1 and BC:ns2 as it allows them to decided
value $1$ in as few as $2$ messages steps.
The coin is then generated as normal in following rounds.

\paragraph{Optimize termination.}
While non-faulty processes in the signature based algorithms are able to terminate immediately after broadcasting a proof
of decision, in the algorithms that do not use signatures non-faulty processes execute one or more additional rounds to ensure
other non-faulty processes decide.
As an optimization, in order to avoid executing unnecessary rounds, a non-faulty process that decides in round
$r$ with binary value $b$ will only execute round $r+1$ if $\neg b$ is valid in round $r$
(i.e. this indicates that a non-faulty process may not have decided in round $r$).

\section{Combination algorithms.}

In this section two new algorithms are presented that simply combine certain parts of the previously described
algorithms.

\subsection{BC:s3}
Algorithm BC:s3 is a combination of BC:s1 and BC:s2.
The first two rounds of BC:s3 are equivalent to the first two rounds of BC:s1 when the value of the coin is predefined as
$1$ for round $1$ and $0$ for round $2$.
Rounds $3$ and later are equivalent to the rounds of BC:s2.
Additionally, proofs of validity from rounds $1$ and $2$ use the construction from BC:s1 when used in
later rounds.

The goal of this algorithm is to use the lower message complexity of algorithm BC:s1 in case the system
can terminate quickly, then to transfer to the algorithm of BC:s2 that can decide either $0$ or $1$.
As BC:s1 can only decide the same value of the coin in a round this avoids long tail latencies where the
coin in BC:s1 repeatably flips the opposite of the remaining valid binary value.

\fbox{\parbox{\textwidth}{
        Algorithm BC:s3 properties.
\begin{smallenum}
\item {\bf Consensus message broadcasts per round:} $2$ in the first round, $1$ in the second round, $2$ in all following rounds.
\item {\bf Fastest decision:} After $2$ consensus message broadcasts.
\item {\bf Binary values that can be decided in a round:} $1$ in round $1$, $0$ in round $2$, either $0$ or $1$ in following rounds.
\item {\bf If the first non-faulty process decides binary value $b$ in round $r$, all non-faulty processes decide at the latest:} Either of the following
  (whichever happens first): (i) if $r = 1$ then in round $3$, otherwise in round $r+1$, or
  (ii) when a proof of decision is received from another process.
\item {\bf Signatures:} Required for consensus messages.
\item {\bf Consensus message contents:} Signature, round number, message type identifier, a constant number of bits, and a proof of validity.
\item {\bf Proofs of validity:} Either $(t+1)$ or $(n-t)$ signatures from a previous round, or if using threshold signatures,
  a single threshold signature.  Twice this if the message contains value $\bot$.
\end{smallenum}
}}

\subsection{BC:ns3}
Algorithm BC:ns3 is a combination of BC:ns1 and BC:ns2.
It follows a similar design to that of BC:s3 except for the non-signature based algorithms.
Namely, the first two rounds of BC:ns3 are equivalent to the first two rounds of BC:ns1 when the value of the coin is predefined as
$1$ for round $1$ and $0$ for round $2$, while rounds $3$ and later are equivalent to the rounds of BC:ns2.

Like BC:s3, the goal of BC:ns3 is to the lower message complexity of algorithm BC:ns1 in case the system
can terminate quickly, then to transfer to the algorithm of BC:ns2 that can decide either $0$ or $1$
in order to avoid avoid the long tail latencies where the
coin in BC:ns1 repeatably flips the opposite remaining valid binary value.

\fbox{\parbox{\textwidth}{
        Algorithm BC:ns3 properties.
\begin{smallenum}
\item {\bf Consensus message broadcasts per round:} $2$ to $3$ in round $1$, $1$ to $2$ in round $2$, $4$ to $5$ in following rounds.
\item {\bf Fastest decision:} After $2$ consensus message broadcasts.
\item {\bf Binary values that can be decided in a round:} $1$ in round $1$, $0$ in round $2$, either $0$ or $1$ in following rounds.
\item {\bf If the first non-faulty process decides binary value $b$ in round $r$, all non-faulty processes decide at the latest:} if $r = 1$ then in round $3$,
  otherwise in round $r+1$.
\item {\bf Weak coin supported:} Yes.
\item {\bf Coin threshold:} $(t+1)$.
\item {\bf Signatures:} Not required for consensus messages.
\item {\bf Consensus message contents:} Round number, message type identifier, and a constant number of bits.
\item {\bf Proofs of validity:} Not required.
\end{smallenum}
}}



\section{Experiments}

\subsection{Experiment Configuration.}
This section describes the configuration of the experiments.

\paragraph{Implementation.}

The algorithms are implemented using the Go programming language~\cite{GO} version 1.13.5.
Threshold BLS signatures, EDDSA signatures, and non-interactive proofs of equality of discrete logarithms,
use the Kyber library~\cite{kyber}. Note that this is an experimental library and may not provide the most efficient implementations.
For messages that are not signed
point to point channels are implemented using encryption and authentication.
For this, a Diffie-Hellman type key exchange generates a shared secret which
is used to encrypt and authenticate messages using the NaCl secretbox functions~\cite{nacl}.
Nodes are connected to each other through TCP connections.
Reliable channels are implemented using message retransmission within the consensus implementation (note that the network
used in the experiments was reliable enough that the TCP connections never dropped
and application level retransmission was not required).

\paragraph{Hardware.}
All experiments are run on Google cloud compute platform using \emph{n1-standard-2} virtual machine instances (nodes)
with 2 vCPUs (hardware Hyper-threads on any one of the Skylake, Broadwell, Haswell, Sandy Bridge, or Ivy Bridge Intel CPU platforms),
7.50 GB memory, local SSDs, and up to 10 Gbps network egress bandwidth.
Nodes run the default Debian-10 operating system image provided by Google cloud.

\paragraph{Experiment settings.}
Within each experiment 110 instances of binary consensus are run one after the other, with the first $10$ being ``warm-up'' instances
and the results are taken as the average of the last $100$ consensus instances.
Within each consensus instance a non-faulty node chooses a random binary value as its initial proposal, where depending
on the experiment configuration, a node will choose $1$ with probability $1/3$, $1/2$, or $2/3$.
Note that each node uses a local random value meaning that a consensus instance will not have exactly $1/3$rd, $1/2$,
or $2/3$rds of nodes proposing $1$, but will have this approximately on average over the $100$ consensus instances.

To reduce the variance between experiments, each experiment uses the same random seeds for choosing proposals
(i.e. if node $n$ proposes binary value $b$ in consensus instance $c$ for experiment $e$, then in experiment $e' /neq e$ node $n$ will also
propose $b$ in consensus instance $c$). Furthermore, the random binary values output by the coin flips are also seeded
so that in the same rounds in different experiments have the same coin output (i.e. if the output of the coin in round $r$,
consensus instance $c$, is binary value $b$ in experiment $e$, then in experiment $e' \neq e$ the output of the coin in round $r$,
consensus instance $c$ is also binary value $b$).
Note that given the experiments are run on shared machines and networks on public clouds, there will be still a fair amount
of variance, but this is also perhaps a realistic setting given a limited budget.

\paragraph{Figures.}
The results from each experiment are presented in figures each containing a set of $4$ graphs.
In each figure the top left graph presents the average time in milliseconds to complete a consensus instance
from proposal to decision.
The top right graph shows the average number of kilobytes sent per consensus instance per node.
The bottom left graph shows the average number of messages sent per consensus instance per node.
The bottom right graph shows the average number of rounds executed until decision per consensus instance per node.
The error bars show the minimum and maximum number of rounds executed until decision in a consensus instance for
all nodes.

Note that in the algorithms when a node calls broadcast it sends the message to all nodes,
including itself, but in the results the messages that a node sends directly to itself are not counted
as they are not sent over the network.

\subsection{Algorithm implementation details.}

\paragraph{Concurrency.}
To take advantage of multi-core hardware, the implementations are multi-threaded.
This is mainly to allow cryptographic operations to take place in parallel such as signing, validation,
encryption, decryption, as well as message serialization and deserialization.

\paragraph{Coin CI:tc.}
Coin algorithm CI:tc is implemented using threshold-BLS signatures. The output of the coin is taken as the
first bit output by the BLAKE2B~\cite{blake} cryptographic has function using the combined threshold signature of the coin shares.
A coin share is a signed message as described in Section~\ref{sec:coin} by a node's threshold key share.

\paragraph{Coin CI:pc.}
Coin algorithm CI:pc is implemented using BLAKE2XB, SHA-256, and the Ed25519 curve functions included in the Kyber library.
Like before the value of the coin is taken as the first bit output by cryptographic has function using the combined coin shares.

\paragraph{Coin CI:tce and CI:pce.}
Coin implementations CI:tce and CI:pce are the implementation of coin CI:tc and CI:pc with an additional $\coinecho$
message that is broadcast before the coin as described in Section~\ref{sec:coin}.
This allows the coin to use a $t+1$ threshold in algorithms that would normally use a $n-t$ threshold.
The $\coinecho$ message is encrypted and not signed.

\paragraph{Coin presets.}
If coin presets are used then the first two outputs of the coin are predetermined
as $1$ and $0$ as described in Section~\ref{sec:opt}. No messages are broadcast for the coin in these rounds.

\paragraph{Combining coin and consensus messages.}
As described in Section~\ref{sec:opt} the signature based algorithms support sending the coin message of a round $r$ with the first
consensus message of round $r+1$.
Experiments that use this setting will use the notation CM:Y (combine messages, yes).

\paragraph{Decision in previous round.}
A process may finish executing a round $r$ without deciding and start round $r+1$.
The process may then receive additional messages from round $r$ (arriving late due to network latency)
causing the process to decide in round $r$.
Now depending on the termination conditions the process may no longer need
to participate in round $r+1$.
This may result in experiments where, for example, an algorithm sending more messages
does not necessarily result in a longer time to decide.

\paragraph{Logging, termination, and garbage collection.}
Termination of consensus instances is implemented as described in Section~\ref{sec:alg}.
Furthermore, the state of each instance of consensus including all received
messages is logged to disk, allowing the state to be recovered and replayed in case
of faulty or slow nodes.
By default the last $10$ decided consensus instances are kept in memory, though this value
can be configured.
The disk log is assumed to be infinite and is never trimmed.

\paragraph{Binary Consensus BC:s1 defaults.}
The default configuration of BC:s1 uses threshold BLS signatures and all consensus messages include a proof of validity
(i.e. IP:Y or include proofs, yes).
Given that round $0$ consensus messages use a $(t+1)$ threshold, and all other rounds use an $(n-t)$ threshold,
two keys generated at each node, one for each threshold.
Coin CI:tc is used by default using the same $(n-t)$ threshold key used to sign the consensus messages.
Channels are not encrypted by default.
Coin messages are not combined with consensus messages of the next round by default (i.e. CM:N or combine messages, no).

When using coin implementation CI:pc or CI:pce, the algorithm uses EDDSA signatures to sign messages.
As EDDSA signatures are not threshold signatures, the proofs of validity included with the consensus
messages consist of either $(t+1)$ or $(n-t)$ individual signatures as needed by the algorithm.
The idea of these experiments is to test the algorithms without using BLS based cryptography.
Note that when including proofs of validity, nodes will receive the same signature multiple
times, thus, when a node receives a signature that it has already validated it simply
discards the new copy to avoid unnecessary cryptographic computations.

\paragraph{Binary Consensus BC:s2 defaults.}
The default configuration for algorithm BC:s2 is the same as BC:s1.

\paragraph{Binary Consensus BC:s3 defaults.}
The default configuration for algorithm BC:s3 is the same as BC:s1 and BC:s2.

\paragraph{Binary Consensus BC:ns1 defaults.}
In the default configuration of BC:ns1 channels are encrypted/authenticated and consensus messages are not signed.
Coin CI:tc is used by default using a $(n-t)$ threshold-BLS key generated prior to the experiments.

\paragraph{Binary Consensus BC:ns2 defaults.}
In the default configuration of BC:ns2 channels are encrypted/authenticated and consensus messages are not signed.
Coin CI:tc is used by default using a $(t+1)$ threshold-BLS key generated prior to the experiments.

\paragraph{Binary Consensus BC:ns3 defaults.}
The default configuration for algorithm BC:ns2 is the same as BC:ns2.

\paragraph{Default experiment configurations table.}
Table~\ref{tab:alg-defaults} shows the default configurations for the consensus algorithms
in the experiments.
These configurations are used in all experiments unless otherwise mentioned.


\begin{table}[h!]
\begin{tabular}{ |m{1.8cm}|m{1.5cm}|m{1.4cm}|m{1.5cm}|m{1.2cm}|m{1.4cm}|m{1.45cm}|m{1.5cm}|m{1.5cm}|  }
 \hline
 Algorithm (BC) & Coin Implem. (CI) & Coin thresh.  & Combine Coin (CM) & Coin Presets & Include Proofs (IP) & Signature type & Encrypt channels \\
 \hline
  s1 & tc & $(n-t)$ &  no & no & yes & TBLS & no \\
  s2 & tc & $(n-t)$ &  no & no & yes & TBLS & no \\
  s3 & tc & $(n-t)$ &  no & no & yes & TBLS & no \\
  ns1 & tc & $(n-t)$ &  n/a & no & n/a & n/a & yes \\
  ns2 & tc & $(t+1)$ &  n/a & no & n/a & n/a & yes \\
  ns3 & tc & $(t+1)$ &  n/a & no & n/a & n/a & yes \\
  \hline
\end{tabular}
  \caption{Default configurations for consensus algorithms in the experiments.}\label{tab:alg-defaults}
\end{table}

\subsection{Cryptographic costs}

This section presents the costs of some of the cryptographic operations used by the algorithms.
Benchmarks are performed on n1-standard-2 Google cloud compute instances using the Go benchmarking utility.

\begin{table}[h!]
  \begin{tabular}{|c|c|c|c|}
    \hline
    & Msg Size (bytes) & Sign (ms) & Verify (ms) \\
    \hline
    EDDSA & 109 & 0.28 &  0.49 \\
    \hline
    Thresh BLS & 110 & 0.365 & 4.20 \\
    \hline
  \end{tabular}
  \caption{Costs of signing messages using EDDSA and threshold-BLS signatures from the Kyber library~\cite{kyber}
    on n1-standard-2 Google cloud instances.}\label{tab:sig}
\end{table}

Table~\ref{tab:sig} shows the costs of signatures in terms of message size in bytes and computation
time in milliseconds.
Computing the cryptographic hash of the messages is included in the time calculations.
The main difference between the two implementations is the verification time in which threshold-BLS
signatures are nearly ten times more costly.

A signature is approximately 85 bytes in size.
A signed message is approximately 110 bytes in size, this includes the signature, identifying information, and the message
itself.
When including proofs of validity as required by the signature based consensus algorithms, an additional message and signature(s)
is appended to each consensus message, effectively doubling the total size of the message when using threshold signatures.
Otherwise, if threshold signatures are not used then the proof of validity will contain an additional 85 bytes for each of the $(t+1)$ or $(n-t)$ (as
defined by the algorithm) individual signatures.

\begin{table}[h!]
  \begin{tabular}{|c|c|c|c|}
    \hline
    & Msg Size (bytes) & Encrypt (ms) & Decrypt (ms) \\
    \hline
    NaCl SecretBox & 70 & 0.0007 &  0.0017 \\
    \hline
  \end{tabular}
  \caption{Costs of encrypting and decrypting messages (including authenticating)
    using NaCl Secretbox functions from the Golang /crypto/x/ library~\cite{nacl}
    on n1-standard-2 Google cloud instances.}\label{tab:enc}
\end{table}

Table~\ref{tab:enc} shows the costs of encrypting and authenticating messages in terms of message size in bytes and computation time
in milliseconds using the NaCl secretbox Golang /crypto/x/ library~\cite{nacl}.
An encrypted/authenticated message is approximately 70 bytes in size. This includes the message and overhead needed by the encryption
and authentication algorithm.
The computation costs of encryption, authentication and decryption are a small fraction of the cost of signing a message.
Note that the algorithms would remain correct if messages are just authenticated and not encrypted,
this may provide a small performance improvement.

\begin{table}[h!]
  \begin{tabular}{|c|c|c|c|}
    \hline
    & Msg size (bytes) & Share gen (ms) & Share Verify (ms) \\
    \hline
    CI:pc & 212 & 1.40 & 1.60 \\
    \hline
    CI:tc & 110 & 0.365 & 4.20 \\
    \hline
  \end{tabular}
  \caption{Size of coin shares and time to generate and verify a share for coin algorithms CI:pc and CI:tc
  implemented using the Kyber library on n1-standard-2 Google cloud instances.}\label{tab:coin}
\end{table}

Table~\ref{tab:coin} shows the overhead of coin shares in terms of message size in bytes and computation time in milliseconds.
As CI:tc uses threshold signatures, the costs are the same as signing and verifying a message using threshold-BLS signatures.
For CI:pc the size of a coin share (message) which includes the non-interactive proof of discrete logarithm equality is approximately 212 bytes.
While generating a share using CI:pc is more than twice the cost of using CI:tc, the cost of verifying a share is less than half
that of CI:tc.
This is significant given that each process will need to verify $(t+1)$ or $(n-t)$ shares for each coin.

\begin{table}[h!]
  \begin{tabular}{|c|c|c|c|c|c|c|c|}
    \multicolumn{3}{c}{} & \multicolumn{5}{c}{Node count} \\
    \hline
    \multicolumn{2}{|c|}{} & Threshold & 4 & 8 & 16 & 32 & 48 \\
    \hline
    \multirow{4}{*}{CI:pc} & \multirow{2}{*}{Combine} & $(t+1)$ & 0.71 & 1.07 & 2.07 & 3.84 & 5.58 \\
                           & & $(n-t)$ & 1.06 & 1.77 & 3.85 & 7.34 & 11.10 \\
                           & \multirow{2}{*}{Coin gen} & $(t+1)$ & 3.90 & 6.06 & 11.92 & 21.74 & 31.16 \\
                           & & $(n-t)$ & 5.93 & 9.93 & 21.90 & 41.01 & 61.47 \\
    \hline
    \multirow{4}{*}{CI:tc} & \multirow{2}{*}{Combine} & $(t+1)$ & 0.21 & .22 & 0.60 & 1.03 & 1.71 \\
                           & & $(n-t)$ & 0.22 & 0.41 & 1.05 & 2.28 & 3.63 \\
                           & \multirow{2}{*}{Coin gen.} & $(t+1)$ & 8.64 & 12.93 & 26.09 & 48.38 & 69.92 \\
                           & & $(n-t)$ & 13.36 & 21.67 & 47.21 & 91.18 & 134.52 \\
    \hline
  \end{tabular}
  \caption{Computation cost in milliseconds
    of combing a threshold of coin shares and generating the output of the coin
    for algorithms CI:pc and CI:tc using different node counts
    implemented using the Kyber library on n1-standard-2 Google cloud instances.}\label{tab:cointhrsh}
\end{table}

Table~\ref{tab:cointhrsh} shows the cost of computing the coin value for different thresholds and different numbers
of nodes.
The value \emph{Combine} represents the costs to combine the threshold of shares together.
The value \emph{Coin gen.} represents the cost of all the operations needed to compute the random output of the coin
from a set of threshold shares.
This includes validating the shares, combining them, and hashing the output.
Note that this value represents the total time of performing these operations sequentially.
In the implementations of the consensus algorithms the validation of shares is computed in parallel.
As expected, the principal cost comes from the share validation, meaning that using the smaller $t+1$
threshold and the CI:pc implementation results in the best performance.

\subsection{Single region experiments.}

Figures~\ref{fig:s-l}-\ref{fig:ns-ce-l} presents experiments using $4$ n1-standard-2 Google cloud nodes
within the same data-center region.
The round trip latency between nodes is approximately $0.15$ milliseconds.
Given the low latency and high bandwidth available within a region these experiments
are mainly CPU bound.

\subsubsection{Default configurations with and without coin presets.}

\begin{figure*}[ht!]
\begin{center}
    \includegraphics[scale=0.50]{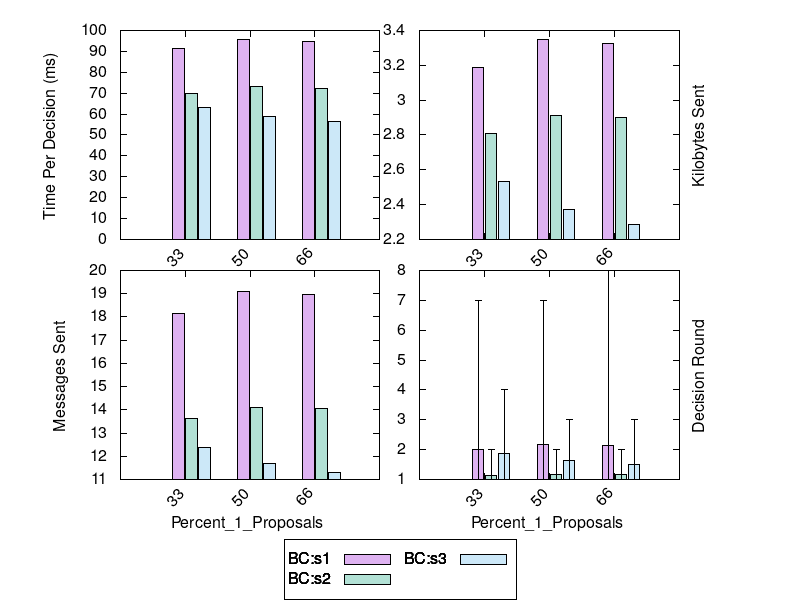}
  \end{center}
  \caption{4 node single region experiment of the signature based algorithms with their default configurations.}
  \label{fig:s-l}
\end{figure*}

Figure~\ref{fig:s-l} shows the results of the experiment using the signature based algorithms
with their default configurations.
The approximate number of nodes proposing $1$ is indicated on the y-axis at $33$, $50$, and $66$ percent.
The average duration of a consensus instance ranges from around $50$ to $100$ milliseconds,
which is many times higher than the network latency.
Furthermore, the number of bytes transferred,
between 2.2 to 3.4 kilobytes, is tiny compared to the available network throughput.
Here the overhead is largely coming from the cryptography.
A direct relation is seen between latency and number of messages sent, where higher latency
is indicated by a higher message count due to additional signature validation.

On average BC:s1 is terminating after approximately $2$ rounds, BC:s2 is deciding after
slightly more than $1$ round on average, and BC:s3 somewhere in-between round $1$ and $2$.
When deciding in round $2$, a node running BC:s1 will have broadcast $5$ messages,
the initial proposal, followed by two rounds consisting of an $\auxm$ message and
a coin share. For BC:s2, when deciding in round $1$ a node will have broadcast $3$ messages,
the initial proposal, followed by a $\prevote$ and a $\mainvote$ message.
For BC:s2, deciding in round $1$ will consist of $2$ message broadcasts,
a broadcast of the initial proposal followed
by the broadcast of an $\auxm$ message, and deciding in round $2$ will add an
additional $\auxm$ message broadcast.
This shows a direct coloration between message broadcasts and consensus duration,
with BC:s1 having the longest duration, and BC:s3 having the shortest.
Furthermore, an important observation is that in algorithms BC:s2 and BC:s3 nodes are frequently
deciding before they compute a single coin value.
This is true even with the highest level of initial disagreement where half the nodes
are proposing $1$.
Note that an adversarial network could generate schedules that force nodes to always generate
at least $1$ coin when there is initial disagreement, but here, given that the network
is well behaved this is not the case.

In all cases BC:s1 has the highest maximum decision round
at either $5$ or $6$.
This comes from the fact that BC:s1 can only decide the value of the coin in each round,
and if only a single binary value is valid it may take several rounds before the coin outputs
that value.
This motivates the use of algorithm BC:s3 which uses the smaller message count of BC:s1
to decide quickly in the first or second round when possible, then switching to the construction
of BC:s2 for later round in order to attempt to avoid high maximum round count.

\begin{figure*}[ht!]
\begin{center}
    \includegraphics[scale=0.50]{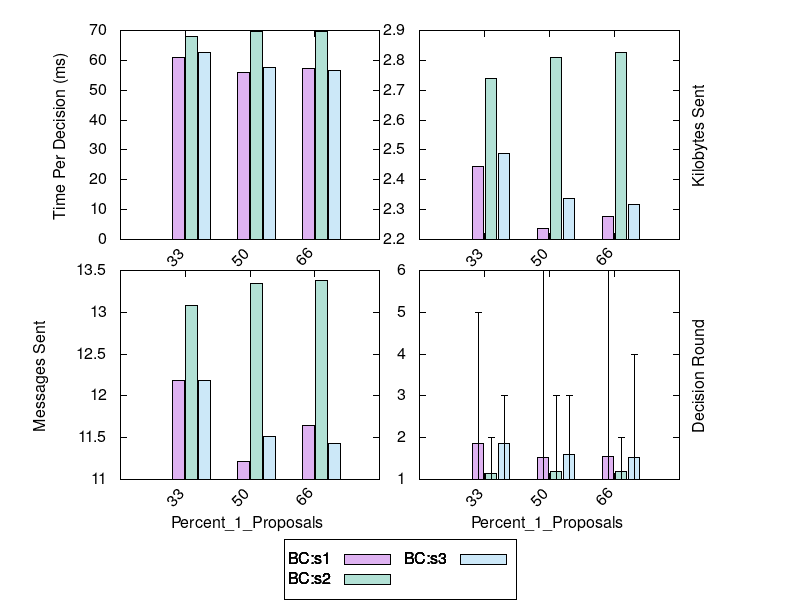}
  \end{center}
  \caption{4 node single region experiment of the signature based algorithms using coin presets.}
  \label{fig:s-p-l}
\end{figure*}

Figure~\ref{fig:s-p-l} shows the results of the same experiment except when using coin presets.
For each consensus instance, the first output of the coin is predefined as $1$ and the second output is $0$.

Here the performance of BC:s1 becomes much better than without coin presets (Figure~\ref{fig:s-l})
and now performs similarly to BC:s3. The reason for this is that by using coin presets, algorithm BC:s1
can terminate in many cases without computing a threshold coin value, which was already
the case for the other algorithms. Still, a malicious network scheduler could prevent this
and force all algorithms to compute coin values.

Algorithms BC:s1 and BC:s2 perform similarly to as they did without
using coin presets, as in both cases the algorithms rarely reach the point where they
must generate a threshold coin.
Like in the previous experiment, algorithm BC:s1 has the highest maximum decision round.
Even though on average the algorithm performs well, it has high variance between consensus
instances.
Its maximum decision round of $6$ uses $11$ message broadcasts, while the maximum
decision round of $3$ for BC:s2 uses $7$ message broadcasts, and the maximum
decision round of $4$ for BC:s3 also uses $7$ message broadcasts.

\begin{figure*}[ht!]
\begin{center}
    \includegraphics[scale=0.50]{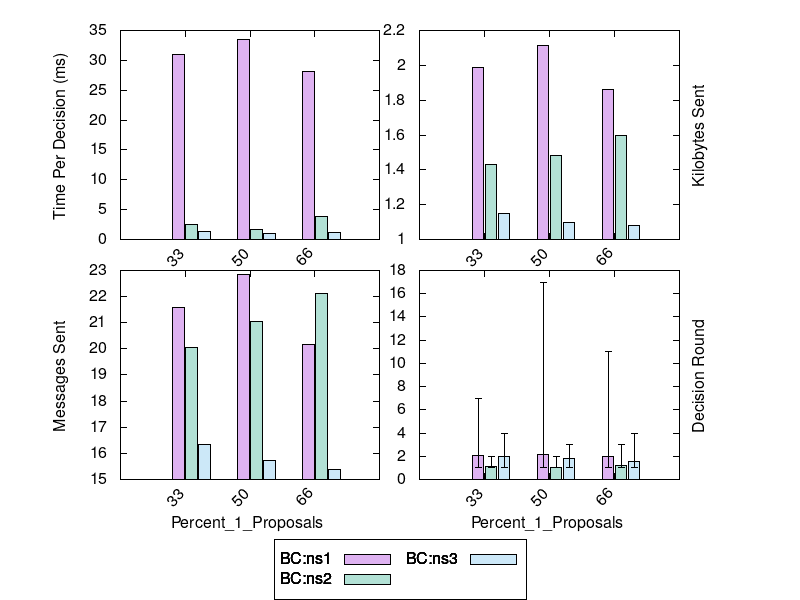}
  \end{center}
  \caption{4 node single region experiment of the non-signature based algorithms using their default configurations.}
  \label{fig:ns-l}
\end{figure*}

Figure~\ref{fig:ns-l} shows the results of the experiment using the non-signature based algorithms
using their default configurations with the amount of nodes proposing $0$ or $1$ shown on the y-axis.
The decision time of the non-signature algorithms is much lower than that of the signature
algorithms. This is due to signing messages being much more expensive than
encrypting them. Furthermore, even though the number of messages sent is higher than the
signature based algorithms ($15$ to $23$ versus $11$ to $19$), the number of kilobytes sent is lower
($2.2$ to $3.4$ versus $1$ to $2.2$) due to the additional bytes needed by the signatures.

Between the non-signature consensus algorithms BC:ns1 has much higher time per decision (around $30$ milliseconds)
than BC:ns2 (around $2$ milliseconds) and BC:ns3 (around $1$ millisecond).
The reason for this is similar to the reason why BC:s1 is slower than the other signature based algorithms,
as BC:ns1 is computing on average $2$ threshold coins per consensus instance while
BC:ns2 and BC:ns3 are not generating any.
Furthermore BC:ns1 has the highest maximum decision round given that it can only decide the value of the coin
in a round.

\begin{figure*}[ht!]
\begin{center}
    \includegraphics[scale=0.50]{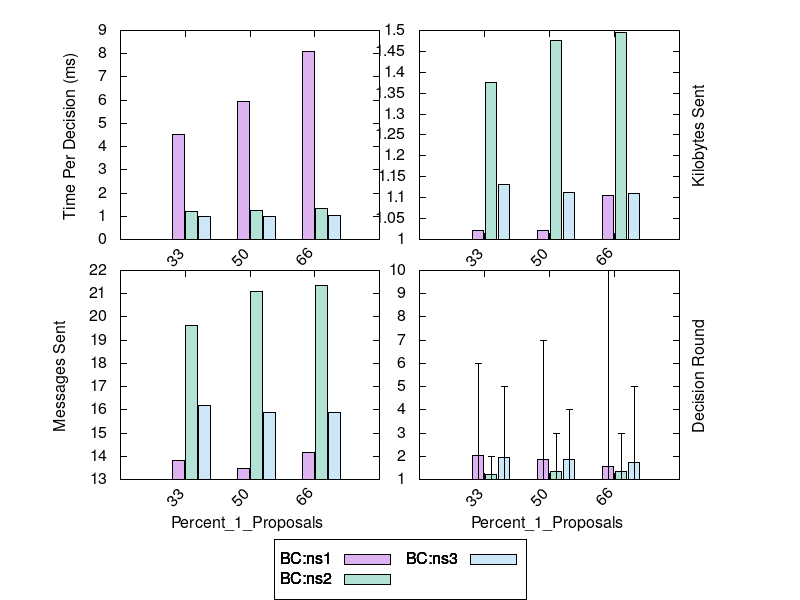}
  \end{center}
  \caption{4 node single region experiment of the non-signature based algorithms with coin presets.}
  \label{fig:ns-p-l}
\end{figure*}

When using coin presets as seen in Figure~\ref{fig:ns-p-l}, the time per decision of BC:ns1 drops to
between $4$ and $8$ milliseconds, while BC:ns1 and BC:ns2 are around $1$ millisecond.
Even though in most cases BC:ns1 decides on round $2$ without computing a threshold coin value,
it still has a higher maximum decision round than the other algorithms, where the cost of computing
a coin is so high that the average decision time is increased.
Overall BC:ns3 has the most efficient performance, taking advantage of the efficient message
complexity of BC:ns1 in most cases without having a high maximum decision round.

\subsubsection{With threshold common coin CI:pc.}
In this section experiments are run using threshold common coin CI:pc.

\begin{figure*}[ht!]
\begin{center}
    \includegraphics[scale=0.50]{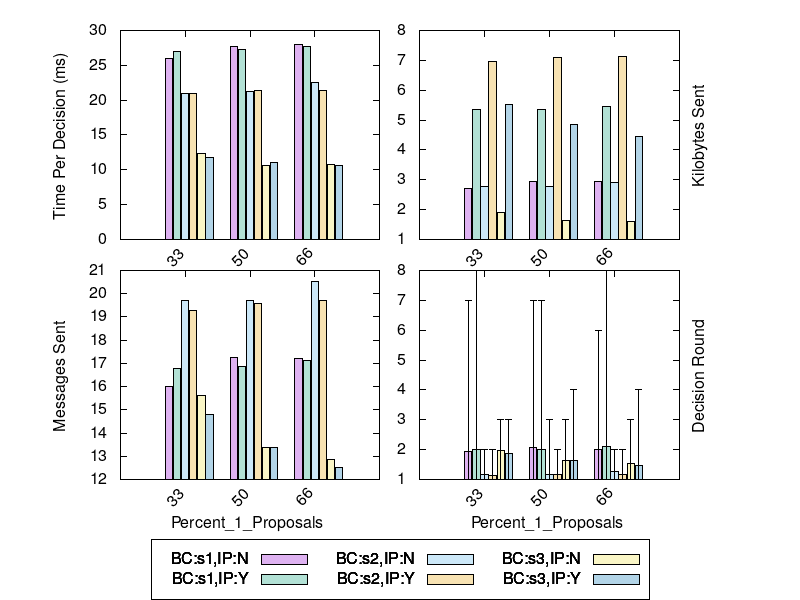}
  \end{center}
  \caption{4 node single region experiment of the signature based algorithms using threshold common coin CI:pc,
    where IP:Y (IP:N) means consensus message do (not) include proofs of validity.}
  \label{fig:s-c2-l}
\end{figure*}

Figure~\ref{fig:s-c2-l} shows the results of the experiment using the signature based algorithms
with the threshold common coin CI:pc where IP:N means consensus message do not include proofs of validity
and IN:Y means they are included.
Note that as mentioned in the experiment settings, when using coin CI:pc consensus messages are signed using
EDDSA keys, therefore proofs of validity contain $(t+1)$ or $(n-t)$ individual signatures.

In these experiments the decision time is much lower than when using threshold-BLS signatures
due to threshold-BLS cryptography being more computationally expensive.
The decision time drops from $90$ to $25$ milliseconds for BC:s1, from $70$ to $20$ for BC:s2 and from $50$
to $10$ for BC:s3.
Whether or not proofs of validity are included has minimal impact on decision time, but at least doubles the number
of bytes sent.
In any case, the number of bytes sent is at most 7 kilobytes which is nowhere near the maximum 10Gbps throughput.

\begin{figure*}[ht!]
\begin{center}
    \includegraphics[scale=0.50]{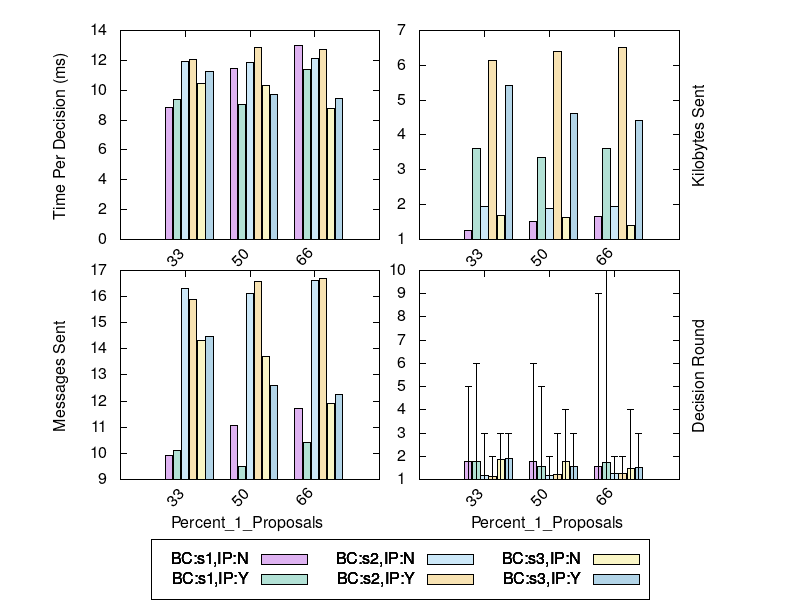}
  \end{center}
  \caption{4 node single region experiment of the signature based algorithms with coin presets using common coin CI:pc.}
  \label{fig:s-p-c2-l}
\end{figure*}

Figure~\ref{fig:s-p-c2-l} shows the results of the experiment using the signature based algorithms
with threshold common coin CI:pc with coin presets.
In this experiment, the decision time drops below $14$ milliseconds in all cases, with BC:s1 and BC:s3
having decision time as low as $8$ milliseconds.
While this is still much higher than the non-signature based algorithms, it is much faster than when
using threshold-BLS signatures.

\begin{figure*}[ht!]
\begin{center}
    \includegraphics[scale=0.50]{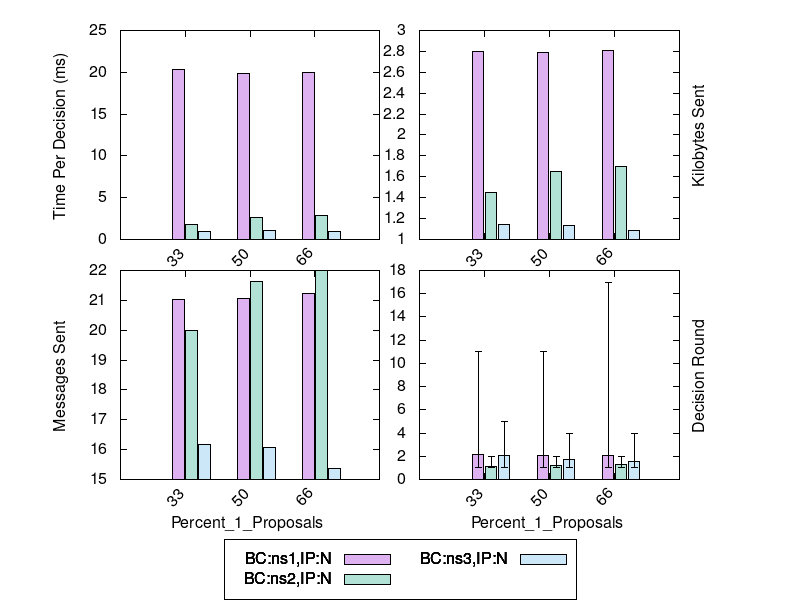}
  \end{center}
  \caption{4 node single region experiment of the non-signature based algorithms using common coin CI:pc.}
  \label{fig:ns-c2-l}
\end{figure*}

Figure~\ref{fig:ns-c2-l} shows the results of the experiment using the non-signature based algorithms with
threshold common coin CI:pc.
Given that BC:ns2 and BC:ns3 rarely need to compute the value of the coin, they preform similarly
to when using coin CI:tc.
The decision time of BC:ns1 drops to around $20$ milliseconds versus around $30$ when using coin CI:tc.

\begin{figure*}[ht!]
\begin{center}
    \includegraphics[scale=0.50]{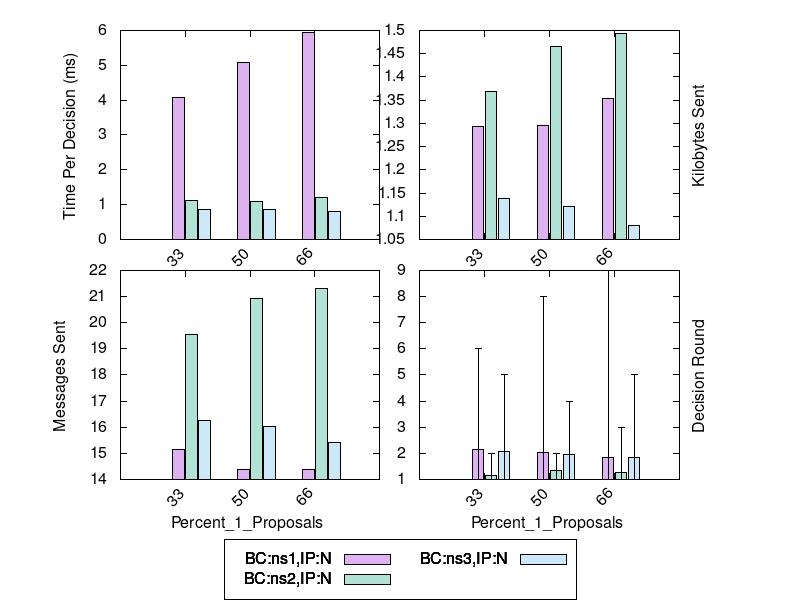}
  \end{center}
  \caption{4 node single region experiment of the non-signature based algorithms with coin presets using common coin CI:pc.}
  \label{fig:ns-p-c2-l}
\end{figure*}

Figure~\ref{fig:ns-p-c2-l} shows the results of the experiment using the non-signature based algorithms with
threshold common coin CI:pc with coin presets.
The main difference here from previous experiments is that the decision time of BC:ns1 drops to between
$4$ and $6$ milliseconds, compared to between $4$ to $8$ milliseconds when using coin CI:tc.
There is not a huge difference here as in most consensus instances the value of the coin is not computed.

\subsubsection{With different coin thresholds.}
\begin{figure*}[ht!]
\begin{center}
    \includegraphics[scale=0.50]{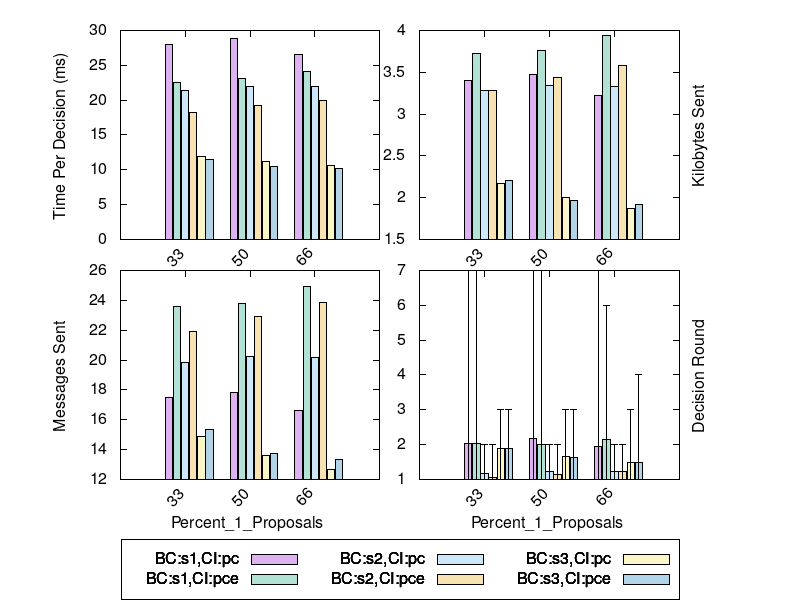}
  \end{center}
  \caption{4 node single region experiment of the signature based algorithms using common coins CI:pc and CI:pce where
    consensus messages do not contain proofs of validity.}
  \label{fig:s-ce-l}
\end{figure*}

\begin{figure*}[ht!]
\begin{center}
    \includegraphics[scale=0.50]{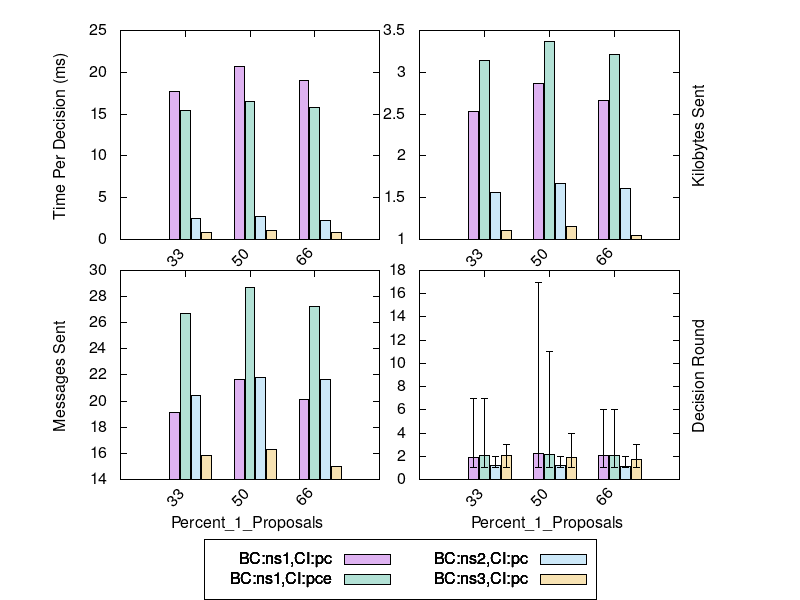}
  \end{center}
  \caption{4 node single region experiment of the non-signature based algorithms using common coins CI:pc and CI:pce.}
  \label{fig:ns-ce-l}
\end{figure*}

Figures~\ref{fig:s-ce-l}-\ref{fig:ns-ce-l} show the results of the experiments when using coin CI:pc and CI:pce.
As described in Section~\ref{sec:alg} coin CI:pc uses an $(n-t)$ threshold, while CI:pce uses a $(t+1)$ threshold,
but requires an extra message broadcast step.

Figure~\ref{fig:s-ce-l} shows the results of the experiment using the signature based algorithms with coin CI:pc and CI:pce.
Consensus messages do not include proofs of validity and coin presets are not used.
All algorithms see a decrease in decision time by using the lower coin threshold as less cryptographic operations are required.
Algorithm BC:s1 has the biggest decrease as it computes the most threshold coin values.
The additional message broadcast of CI:pce increases the number of bytes sent, but the impact is minor given
the latency and the load on the network is very low.

Figure~\ref{fig:ns-ce-l} shows the results of the experiment using the non-signature based algorithms with coin CI:pc and CI:pce.
Coin presets are not used.
Note that only algorithm CI:ns1 uses coin CI:pce as algorithms BC:ns2 and BC:ns3 use a $(t+1)$ coin threshold by default.
Here BC:ns1 sees a small decrease in decision time when using CI:pce due to the fewer cryptographic operations required.

\subsection{Multiple-region experiments}
Figures~\ref{fig:s-us}-\ref{fig:ns-ce-us} presents experiments using $16$ n1-standard-2 Google cloud nodes
across four different regions in central and western United States (Council Bluffs, Iowa, The Dalles, Oregon,
Los Angeles, California and Salt Lake City, Utah).
The round trip latency between nodes within the same region is around $2$ milliseconds
(note that this is higher than the intra-region latency of the previous experiments as those used
local IP addresses, while here public IP addresses are used), with the
inter-region latency being between around $20$ to $35$ milliseconds.

\subsubsection{Default configurations with and without coin presets.}

\begin{figure*}[ht!]
\begin{center}
    \includegraphics[scale=0.50]{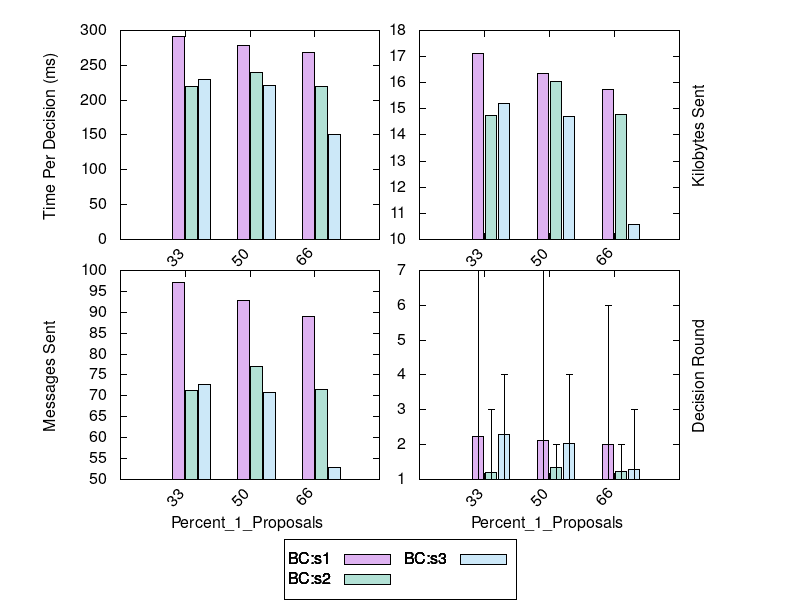}
  \end{center}
  \caption{16 node multi region experiment of the signature based algorithms.}
  \label{fig:s-us}
\end{figure*}

Figure~\ref{fig:s-us} shows the results of the experiments using the signature based algorithms
in their default configurations.
Overall the results follow similar patters as the single region experiments
(Figure~\label{fig:s-us}), except on a larger scale given the higher network latency
between nodes and higher number of nodes.
For example there is a significant consensus decision time difference when using BC:s3 with $33$ percent $1$ proposals at
around $220$ milliseconds and $66$ percent one proposals at around $150$ milliseconds.
This is due to algorithm BC:3 being able to terminate with value $1$ after $2$ message broadcasts versus
value $0$ requiring at least $3$ message broadcasts.
The combination of increased latency and the large increase in number of signature validations needed explain this difference.

\begin{figure*}[ht!]
\begin{center}
    \includegraphics[scale=0.50]{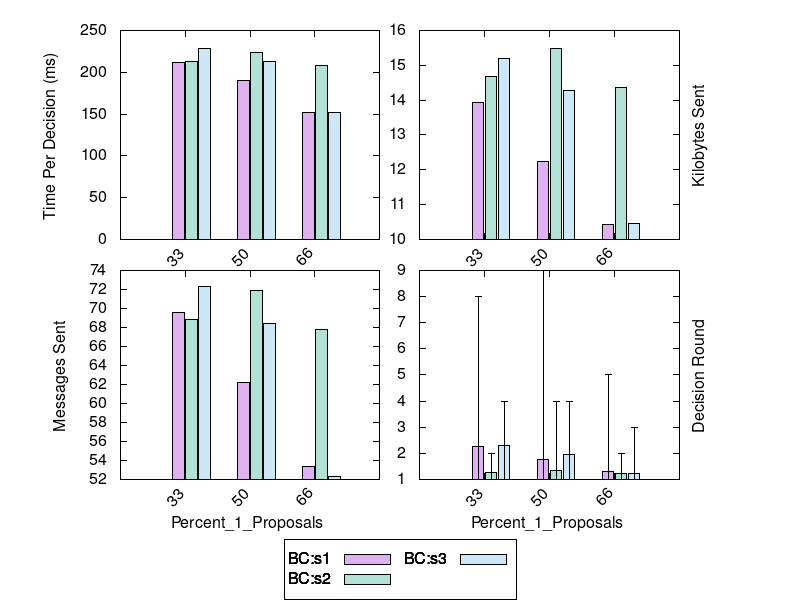}
  \end{center}
  \caption{16 node multi-region experiment of the signature based algorithms (with coin presets).}
  \label{fig:s-p-us}
\end{figure*}

Figure~\ref{fig:s-p-us} shows the results of the experiments using the signature based algorithms
using coin presets.
Again we see a similar patter to the single region experiments where algorithm BC:s1 benefits
the most as in most cases it is now able to terminate without computing a coin value
similar to BC:s2 and BC:s3.
Algorithms BC:s1 and BC:s3 have a significant performance advantage when more nodes start with an initial
proposal of $1$ as they are able to decide after as few as $2$ message broadcasts.

\begin{figure*}[ht!]
\begin{center}
    \includegraphics[scale=0.50]{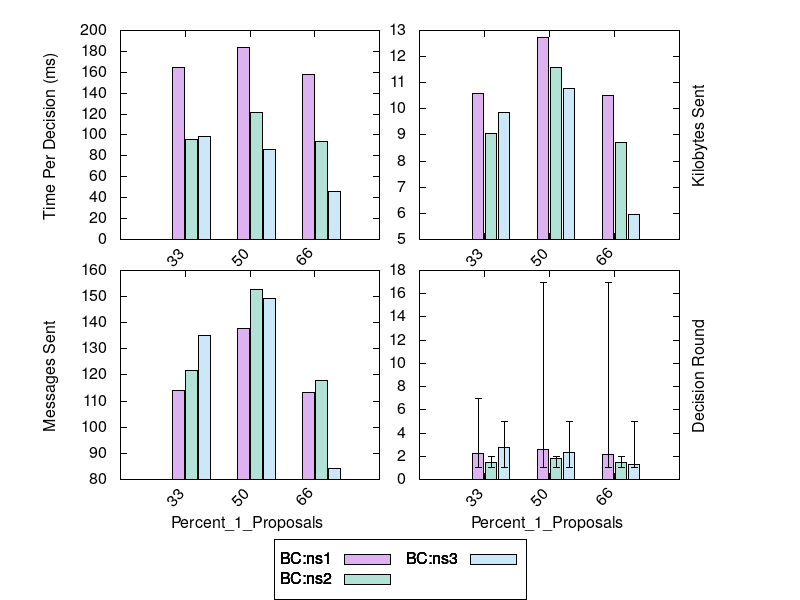}
  \end{center}
  \caption{16 node multi region experiment of the non-signature based algorithms.}
  \label{fig:ns-us}
\end{figure*}

Figure~\ref{fig:ns-us} shows the results of the experiments using the non-signature based algorithms
in their default configurations.
As in the single region experiments, BC:ns1 still has the highest time to decide, owing
mostly to the fact that it ends up having to compute multiple coins values per consensus
instance while the other algorithms do not.

While BC:ns1 broadcasts the fewest messages with $33$ and $50$
percent $1$ proposals, it still has the highest time to decide.
This is for two reasons, firstly, many of the messages are for computing the coin which
results in costly cryptographic operations (note that this is also shown by the fact that
BC:ns1 sends more bytes than the other algorithms as coin messages are larger).
Secondly, in each round the algorithms BC:s2 and BC:s3 perform the broadcast echo message pattern twice
while BC:s1 only uses this patter once per round
(i.e. once $t+1$ messages of the same type are received, that the node broadcasts (echos) the same message
if it has not already).
This echo pattern often takes place in parallel with the broadcasts of other messages so it does not
always result in an increased consensus time.

It is also important to point out that the time per decision is as low as $40$ milliseconds with
BC:ns3 and $66$ percent $1$ proposals, compared to $150$ milliseconds with the signature based BC:s3.
Given that both algorithms can decide in as low as $2$ message broadcast in this setting,
the main additional overhead of BC:s3 is coming from the computational cost of cryptography.

\begin{figure*}[ht!]
\begin{center}
    \includegraphics[scale=0.50]{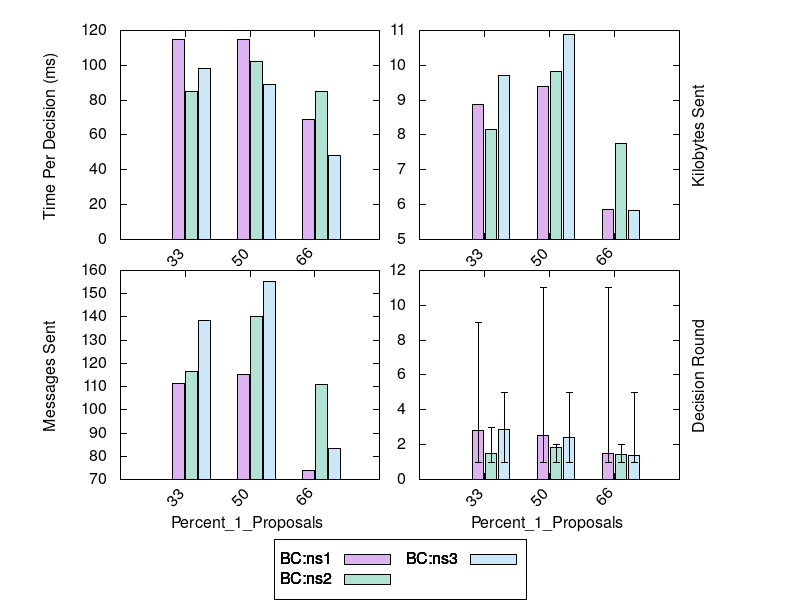}
  \end{center}
  \caption{16 node multi-region experiment of the non-signature based algorithms (with coin presets).}
  \label{fig:ns-p-us}
\end{figure*}

Figure~\ref{fig:s-p-us} shows the results of the experiments using the non-signature based algorithms
using coin presets.
As before this mainly impacts algorithm BC:ns1 as it no longer generates threshold coins
in most consensus instances.
The decision time of BC:ns1 becomes comparable to the other non-signature based algorithms.

\subsubsection{With threshold common coin CI:pc.}

\begin{figure*}[ht!]
\begin{center}
    \includegraphics[scale=0.50]{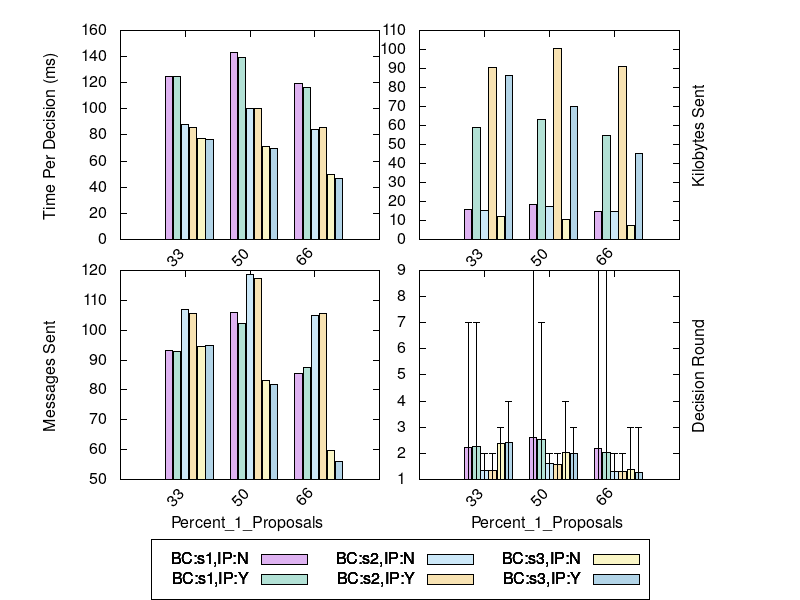}
  \end{center}
  \caption{16 node multi-region experiment of the signature based algorithms using threshold common coin CI:pc
    where IP:Y (IP:N) means consensus message do (not) include proofs of validity.}
  \label{fig:s-c2-us}
\end{figure*}

Figure~\ref{fig:s-c2-us} presents the results of the experiment using the signature based algorithms with threshold
common coin CI:pc where IP:N means consensus messages do not contain proof of validity and IP:Y means they do.
The results here follow similar patterns as when using common coin CI:tc, except with much lower time to decide.
Here the highest average time to decide is around $150$ milliseconds with BC:s1 compared to
$280$ milliseconds with coin BC:s1 and CI:tc.
The minimum average time to decide is around $50$ milliseconds with BC:s3 compared to $150$ milliseconds with coin BC:s3 and CI:tc.
This is directly caused by the lower computational costs of the EDDSA signatures and the CI:pc cryptography.

Including proofs of validity with the consensus message increases the number of bytes significantly, up to as much
as 100 kilobytes per consensus instance per node with BC:s2, but has minimal impact on the time to decide
as it this still fairly low compared to the maximum network throughput.
Given the large difference in the number of bytes
sent with $4$ node experiment as compared to the $16$ node experiment (from a maximum of $6$ kilobytes to $100$ kilobytes),
the network may start becoming a bottleneck when increasing the node count further.
Note that BC:s2 has the highest number of bytes sent partially due to that in certain cases it must include sets of proofs for both $0$
and $1$, whereas BC:s1 consensus messages will contain proofs for a single value.

\begin{figure*}[ht!]
\begin{center}
    \includegraphics[scale=0.50]{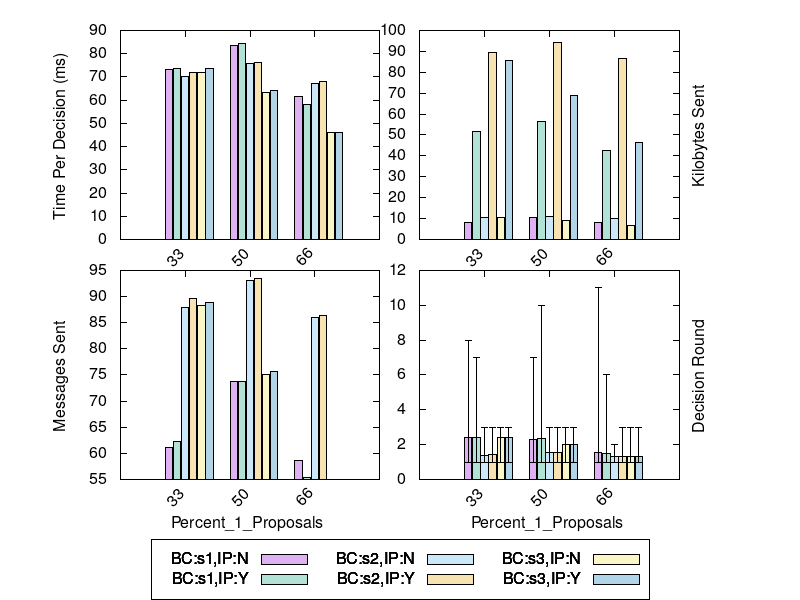}
  \end{center}
  \caption{16 node multi-region experiment of the signature based algorithms with coin presets using common coin CI:pc
        where IP:Y (IP:N) means consensus message do (not) include proofs of validity.}
  \label{fig:s-p-c2-us}
\end{figure*}

Figure~\ref{fig:s-c2-us} presents the results of the experiment using the signature based algorithms with threshold
common coin CI:pc and when using coin presets.
Like in the previous experiments, algorithm BC:s1 benefits the most in this setting,
but BC:s2 and BC:s3 also show a significant performance benefit, as they do not always
terminate in the initial rounds.

\begin{figure*}[ht!]
\begin{center}
    \includegraphics[scale=0.50]{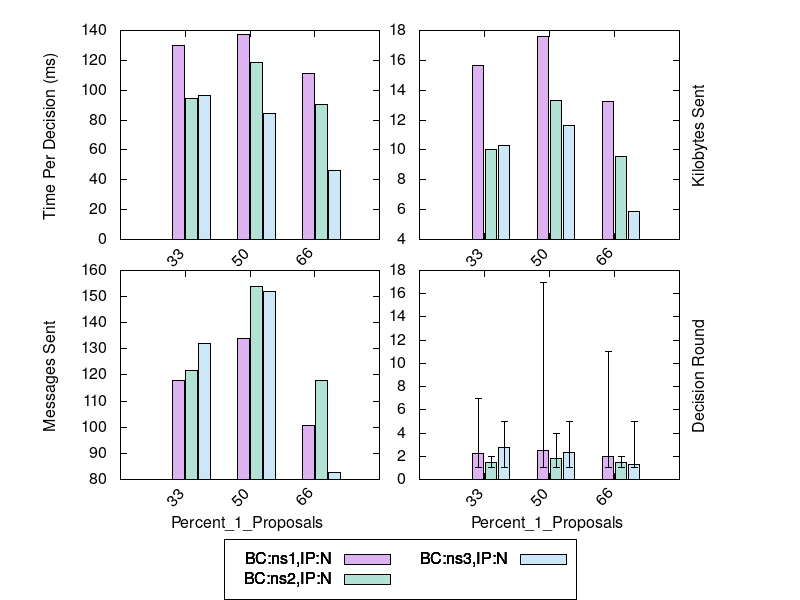}
  \end{center}
  \caption{16 node multi-region experiment of the non-signature based algorithms using common coin CI:pc.}
  \label{fig:ns-c2-us}
\end{figure*}

Figure~\ref{fig:ns-c2-us} presents the results of the experiment using the non-signature based algorithms with threshold
common coin CI:pc.
As in previous experiment without coin presets, BC:ns1 has the largest average time to decide as it must generate
threshold coins in every consensus instance, and BC:ns3 generally having the best performance.

At interesting note is that this is the first experiment configuration where the signature based algorithms end up having
lower time to decide in certain cases when compared to the non-signature algorithms.
In this case the lower number of message steps needed to decide of the signature based algorithms
ends up being more important than the cryptography cost due to higher network latency.
When considering the fastest configurations of BC:s1 and BC:ns1 with $66$ percent $1$ proposals, both algorithms
have an average decision time of around $40$ milliseconds.
The reason is that both these algorithms have a minimum of $2$ message steps to decide.
Given that many reductions to multi-value consensus will use a value of $1$ to indicate a non-$\bot$ value being
decided, this gives an advantage to non-signature based algorithms for also being computationally efficient in this case.

\begin{figure*}[ht!]
\begin{center}
    \includegraphics[scale=0.50]{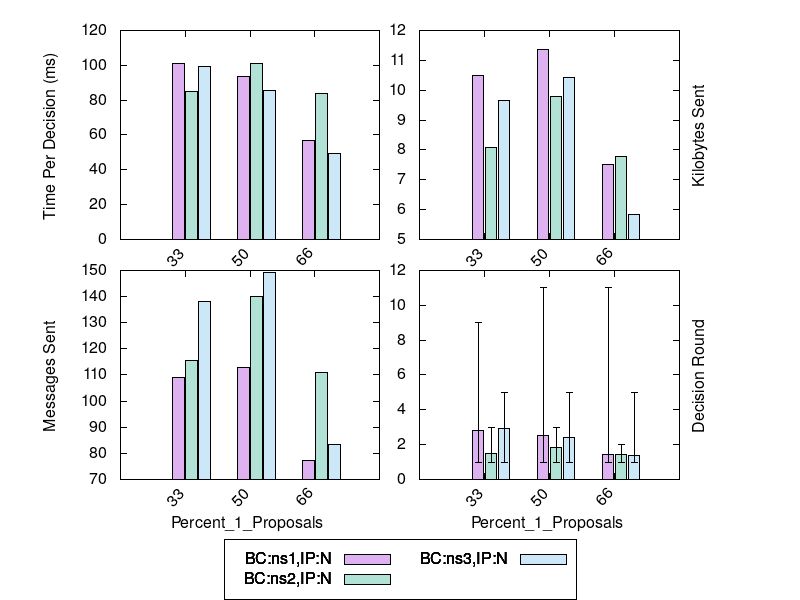}
  \end{center}
  \caption{16 node multi-region experiment of the non-signature based algorithms with coin presets using common coin CI:pc.}
  \label{fig:ns-p-c2-us}
\end{figure*}

Figure~\ref{fig:ns-c2-us} presents the results of the experiment using the non-signature based algorithms with threshold
common coin CI:pc while using coin presets.
As expected using coin presets mainly benefits BC:ns1 due to it not needing to compute the value of threshold coins
in every consensus instance.

\subsubsection{With different coin thresholds.}
\begin{figure*}[ht!]
\begin{center}
    \includegraphics[scale=0.50]{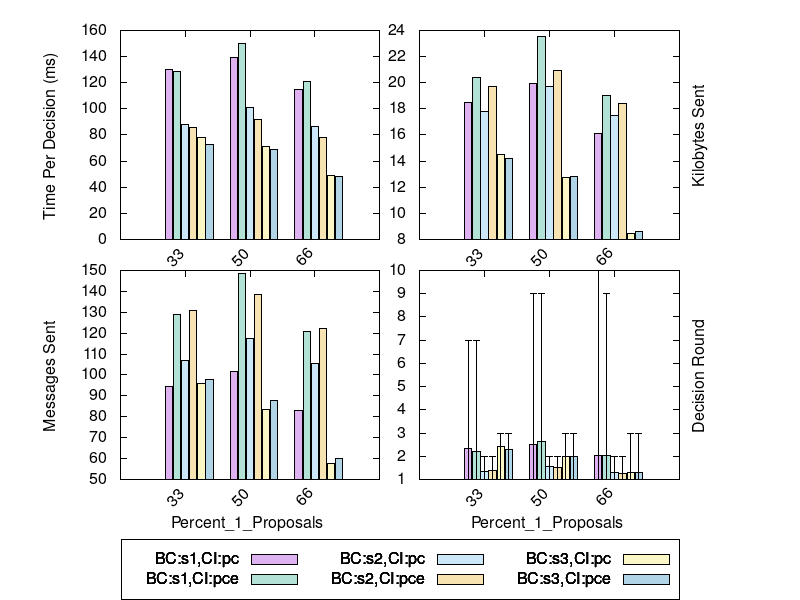}
  \end{center}
  \caption{16 node multi-region experiment of the signature based algorithms using common coins CI:pc and CI:pce where
    consensus messages do not contain proofs of validity.}
  \label{fig:s-ce-us}
\end{figure*}

\begin{figure*}[ht!]
\begin{center}
    \includegraphics[scale=0.50]{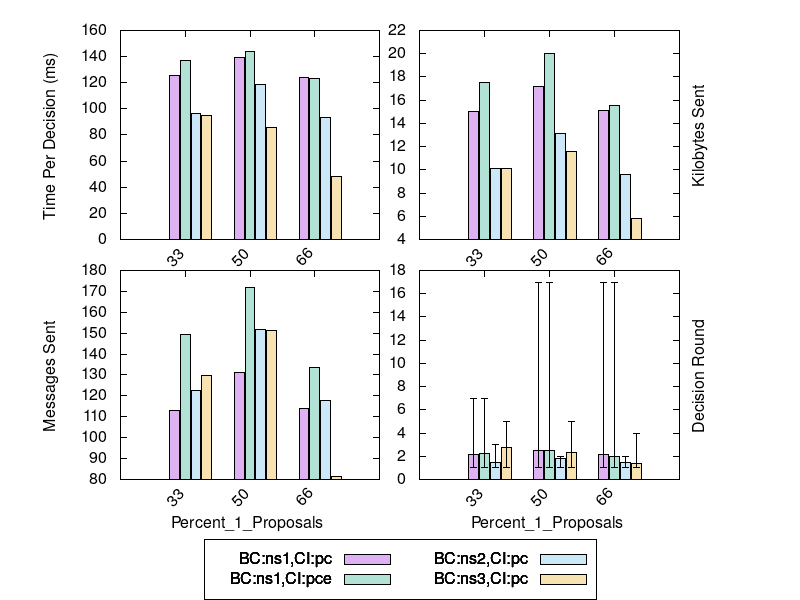}
  \end{center}
  \caption{16 node multi-region experiment of the non-signature based algorithms using common coin CI:pc and CI:pce.}
  \label{fig:ns-ce-us}
\end{figure*}

Figures~\ref{fig:s-ce-us}-\ref{fig:ns-ce-us} shows the results of the experiments when using coin CI:pc and CI:pce.
Coin CI:pc uses an $(n-t)$ threshold, while CI:pce uses a $(t+1)$ threshold,
but requires an extra message broadcast step.

Figure~\ref{fig:s-ce-us} presents the results of the experiment using the signature based algorithms with threshold
common coins CI:pc and CI:pce.
The impact of changing the coin in these experiments does not make a large impact on time to decide,
and mostly just results in additional network overhead.
This is not surprising for BC:s2 and BC:s3 as they are mostly deciding in the first round.
Algorithm BC:s1 has shows a small increase in time to decide with the additional latency caused by the message
broadcast being higher than the reduction in cryptographic computation time.
Overall decision time variations mostly correspond to mostly to variations in average decision round,
likely due to the non-deterministic nature of the experiments.

Figure~\ref{fig:ns-ce-us} presents the results of the experiment using the non-signature based algorithms with threshold
common coin CI:pc and CI:pce.
As before given that BC:ns2 and BC:ns3 already use a $t+1$ coin threshold they are not tested with CI:pce.
For BC:ns1 using CI:pce mostly just adds additional overhead as the reduction in commutation costs due to
the smaller threshold is minimal when compared to the additional latency required for the additional message broadcast.

\subsection{Geo-distributed experiments}
\begin{figure*}[ht!]
\begin{center}
    \includegraphics[scale=0.50]{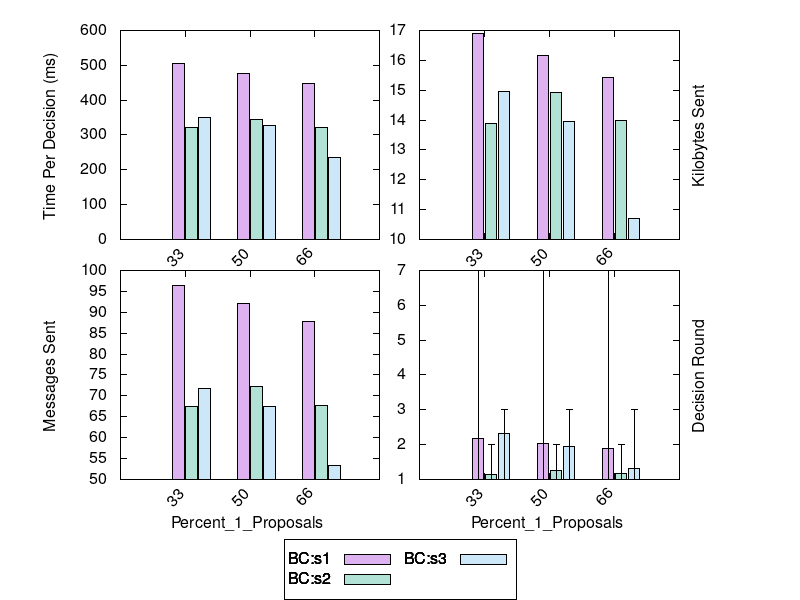}
  \end{center}
  \caption{16 node geo-distributed experiment of the signature based algorithms using default configurations.}
  \label{fig:s-md}
\end{figure*}

\begin{figure*}[ht!]
\begin{center}
    \includegraphics[scale=0.50]{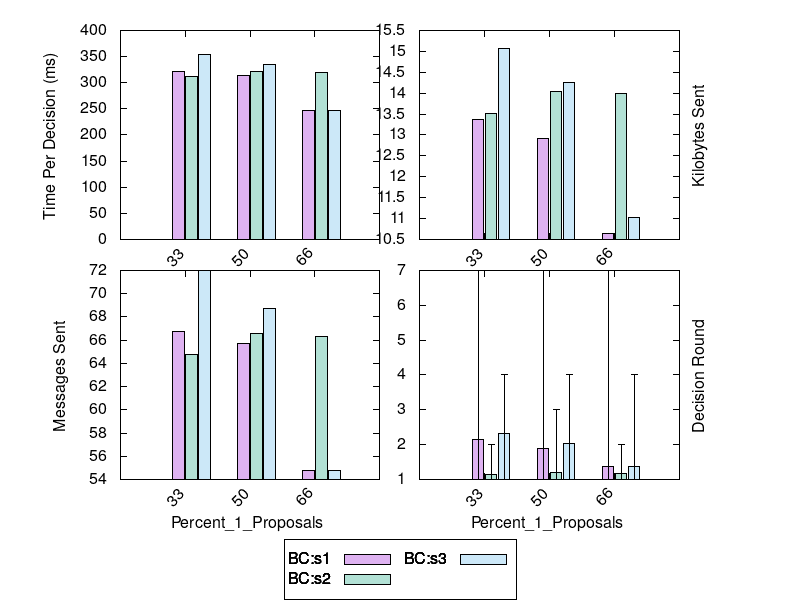}
  \end{center}
  \caption{16 node geo-distributed experiment of the signature based algorithms with coin presets.}
  \label{fig:s-p-md}
\end{figure*}

\begin{figure*}[ht!]
\begin{center}
    \includegraphics[scale=0.50]{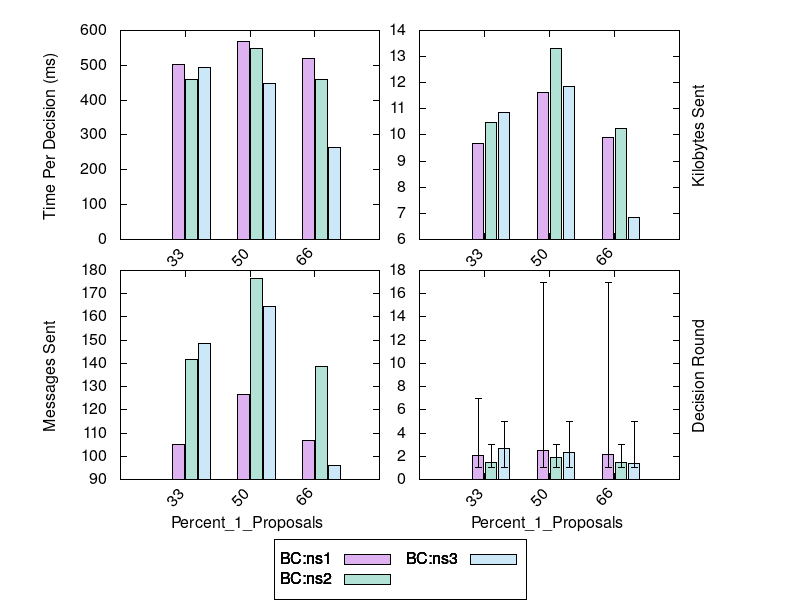}
  \end{center}
  \caption{16 node geo-distributed experiment of the non-signature based algorithms.}
  \label{fig:ns-md}
\end{figure*}

\begin{figure*}[ht!]
\begin{center}
    \includegraphics[scale=0.50]{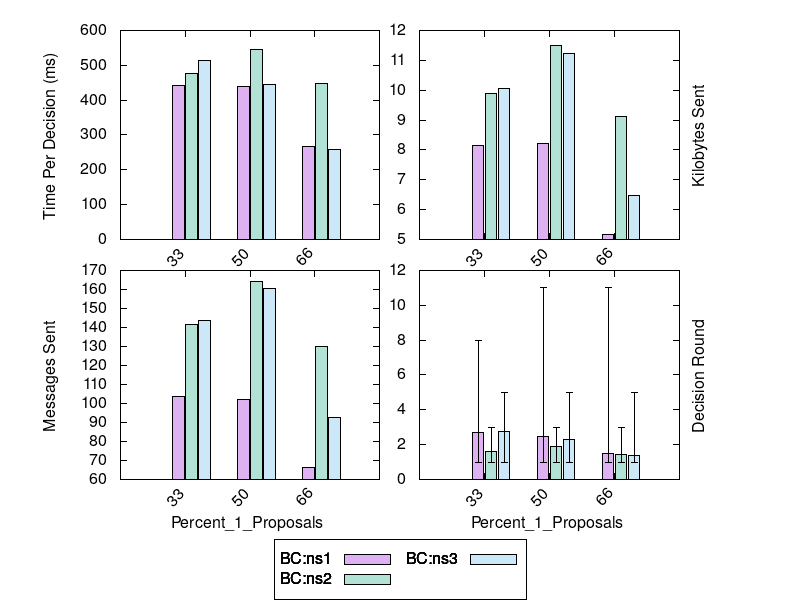}
  \end{center}
  \caption{16 node geo-distributed experiment of the non-signature based algorithms (with coin presets).}
  \label{fig:ns-p-md}
\end{figure*}

Figures~\ref{fig:s-md}-\ref{fig:ns-byz-66} present the results of experiments using $16$ n1-standard-2 Google cloud compute instances
across eight different regions across North America, Europe, Asia and Australia
(Changhua County, Taiwan, Sydney, Australia, Frankfurt, Germany, London, England, Hamina, Finland, Oregon, USA,
Northern Virgina, USA, and Iowa, USA).
The round trip latency between nodes within the same region is around $2$ milliseconds
and inter-region latency ranges from $25$ to $280$ milliseconds.

Figure~\ref{fig:s-md} presents the results of the experiment using the signature based algorithms with
their default configurations.
The results show similar patterns to the previous $16$ node United States based experiment for the same configuration (Figures~\ref{fig:s-us})
except with higher decision times due to the higher network latency between the nodes.
Additionally, given that each message step takes much longer to reach all nodes there is a large difference
between the fastest configuration, BC:s3 at around $250$ milliseconds, and the slowest configuration,
BC:s1 at $500$ milliseconds.

Figure~\ref{fig:s-md} presents the results of the experiment using the signature based algorithms
when using coin presets.
As in previous experiments the coin presets many improve the performance of BC:s1 which now
shows similar performance to BC:s3.

Figure~\ref{fig:ns-md} presents the results of the experiment using the non-signature based algorithms with
their default configurations.
Here the time to decide is higher in most cases than the signature based algorithms due to the additional
messages steps needed and high network latency.
Like that of BC:s3 for the signature based algorithms,
the fastest configuration is BC:ns3 with $66$ percent $1$ proposals is around $250$ milliseconds
as both algorithms are able to terminate in $2$ message steps.

Figure~\ref{fig:ns-md} presents the results of the experiment using the non-signature based algorithms with
coin presets.
As in previous experiments the coin presets benefit mostly BC:ns1, which now performs similarly to BC:ns3.

\subsubsection{Combining messages.}

\begin{figure*}[ht!]
\begin{center}
    \includegraphics[scale=0.50]{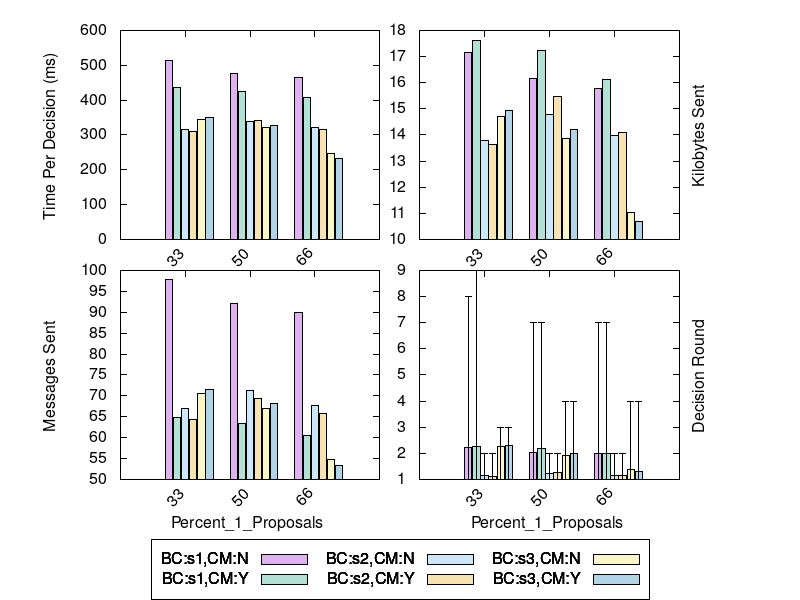}
  \end{center}
  \caption{16 node geo-distributed experiment of the signature based algorithms where the coin messages
    are combined with the first message of the following round.
    CM:N means messages are not combined and CM:Y means messages are combined.}
  \label{fig:s-cm-md}
\end{figure*}

Figure~\ref{fig:s-cm-md} presents the results of the experiment using the signature based algorithm
where the coin messages are combined with the first message of the next round in order to reduce the number
of message steps per round as described in Section~\ref{sec:opt}.
CM:N means messages are not combined and CM:Y means messages are combined.
Note that coin presets are not used as this optimization only provides benefits when the coin is generated.
Algorithms BC:s2 and BC:s3 do not benefit much from this optimization as in most cases they are deciding
before a coin is being computed.
Algorithm BC:s1 shows around a $50$ millisecond improvement in decision time, and a large reduction
in number of messages sent, but with a small increase in number of bytes sent, due to not
always being able to use threshold signatures.
Given that the BC:s1 is seeing benefits even in this experiment where the network is well behaved,
it is expected that all algorithms should benefit similarly in the case of an adversarial network.

\subsubsection{Experiments with faulty behaviors.}
The following faulty (or Byzantine) behaviors are tested, they modify the message broadcast functions as described.

\begin{itemize}
\item {\bf B - Both binary.} Instead of broadcasting a single message with a binary value, multiple messages are broadcast, one for each binary
  value, and also a message with $\bot$ as a value if valid for the given message type.
\item {\bf F - Flip binary.} Binary values in messages are flipped to be the opposite binary value.
  Furthermore, if value of the message can be $\bot$ then $\bot$ is chosen as the value instead.
\item {\bf H - Half/half.} The faulty nodes keep an ordered list of the non-faulty nodes. They then broadcast the normal message
  as given by the algorithm to the front half of the nodes, and a message with the flipped binary value to the back half of the nodes.
\item {\bf HF - Half/half fixed.} The faulty nodes keep an ordered list of the non-faulty nodes. Messages containing the binary value $0$
  are broadcast to the front half of the nodes and the binary value $1$ to the back half of the nodes.
\item {\bf M - Mute.} Faulty nodes broadcast no messages.
\item {\bf N - Non-faulty.} The benchmark is run with no faulty nodes.
\end{itemize}

Note that, given that, in the experiments there are $\numfaults$ faulty nodes, they will neither be able to affect
agreement nor termination, i.e. the non-faulty nodes will still terminate correctly in the expected $O(1)$ rounds.
Furthermore, if the remaining non-faulty nodes all propose the same binary value $b$ then the faulty nodes
cannot effect termination at all as $\neg b$ will never be a valid value to be decided
(by the BC-validity property).

Even though they cannot effect the outcome, the faulty nodes still have some options to effect the performance of
the algorithms.
Consider the following possibilities:
\begin{itemize}
\item Faulty nodes can add additional load to the system, for example to perform a denial of service attack. Although this type
  of attack is not tested here, the \emph{B - Both binary} fault type
  will add extra processing load and network load on the system.
\item Specific to the signature based algorithms, if non-faulty node's messages do not contain proofs of validity, then
  faulty nodes can create situations where certain non-faulty nodes receive messages from other non-faulty nodes that are
  not valid given the state of the node upon reception. In this case the receiver will then have to ask the sender
  for a proof of validity.
  The \emph{H - Half/half} and \emph{H - Half/half fixed} faults can cause this to happen.
  To avoid this, in the experiments all non-faulty nodes include proofs of validity with messages.
\item As previously described, if at least $(n-t)$ non-faulty nodes propose the same binary value $b$ then only that
  value can be decided. This means that in algorithms $s2$, $ns2$, non-faulty nodes will decide in the first round;
  in algorithms $s3$, $ns3$ non-faulty nodes will decided in the first or second round; and
  if coin presets are used in algorithms $s1$ and $ns1$ non-faulty nodes will also decide in the first or second round.
  
  Otherwise, if less than $(n-t)$ non-faulty nodes propose the same value then either $0$ or $1$ can be decided
  and there exist orderings of messages where non-faulty nodes do not decide in the first round(s).
  As seen in the previous experiments with no faults, the algorithms still often terminate in the first round(s)
  even when when approximately $1/2$ of the nodes propose $0$.
  This gives the opportunity for Byzantine nodes to try to prevent this from happening by doing things like
  broadcasting different binary values or the value $\bot$ when possible.
\end{itemize}

\begin{figure*}[ht!]
\begin{center}
    \includegraphics[scale=0.50]{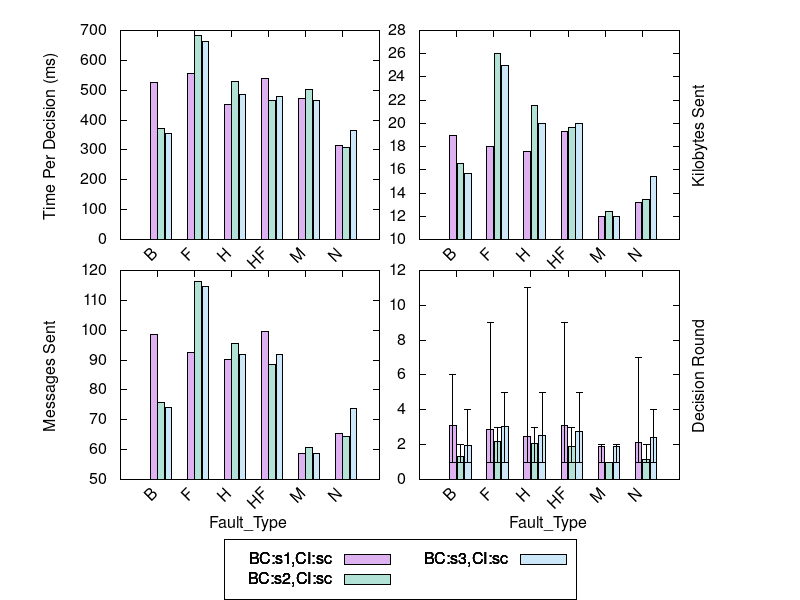}
  \end{center}
  \caption{16 node geo-distributed experiment of the signature based algorithms with coin presets where 1/3 of the nodes exhibit various
    faulty behavior and approximately 1/3 of the non-faulty nodes propose 1.}
  \label{fig:s-byz-33}
\end{figure*}

\begin{figure*}[ht!]
\begin{center}
    \includegraphics[scale=0.50]{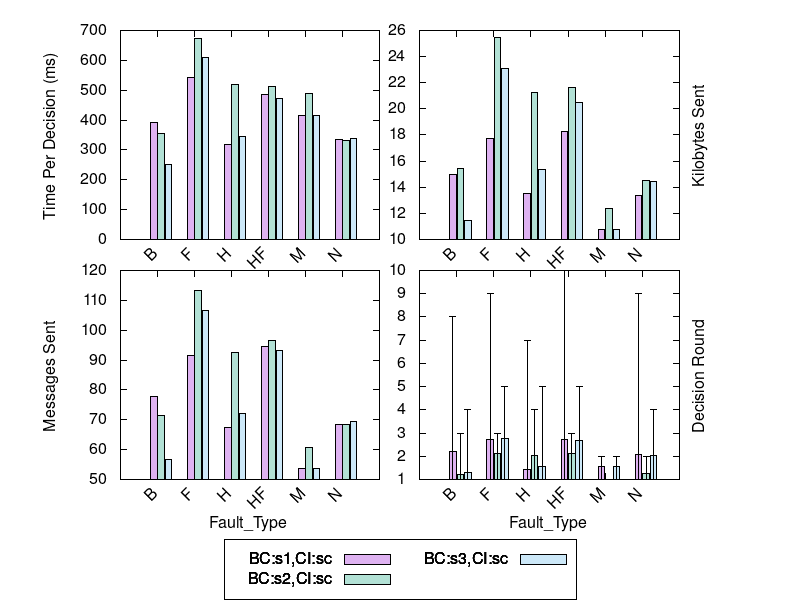}
  \end{center}
  \caption{16 node geo-distributed experiment of the signature based algorithms using coin presets where 1/3 of the nodes exhibit various
    faulty behaviors and approximately 1/2 of the non-faulty nodes
  propose 1.}
  \label{fig:s-byz-50}
\end{figure*}

\begin{figure*}[ht!]
\begin{center}
    \includegraphics[scale=0.50]{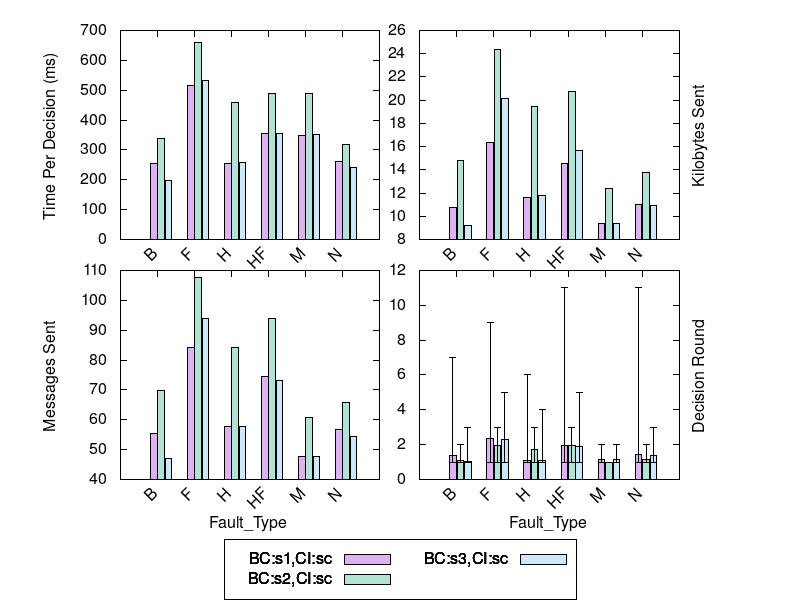}
  \end{center}
  \caption{16 node geo-distributed experiment of the signature based algorithms using coin presets where 1/3 of the nodes exhibit various
    faulty behavior and approximately 2/3 of the non-faulty nodes  propose 1.}
  \label{fig:s-byz-66}
\end{figure*}

\begin{figure*}[ht!]
\begin{center}
    \includegraphics[scale=0.50]{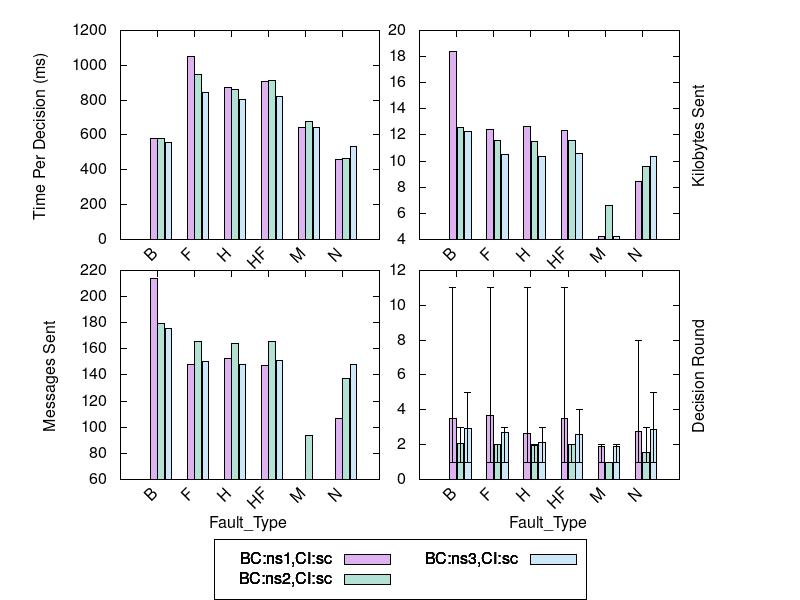}
  \end{center}
  \caption{16 node geo-distributed experiment of the non-signature based algorithms using coin presets where 1/3 of the nodes exhibit various
    faulty behavior and approximately 1/3 of the non-faulty nodes  propose 1.}
  \label{fig:ns-byz-33}
\end{figure*}

\begin{figure*}[ht!]
\begin{center}
    \includegraphics[scale=0.50]{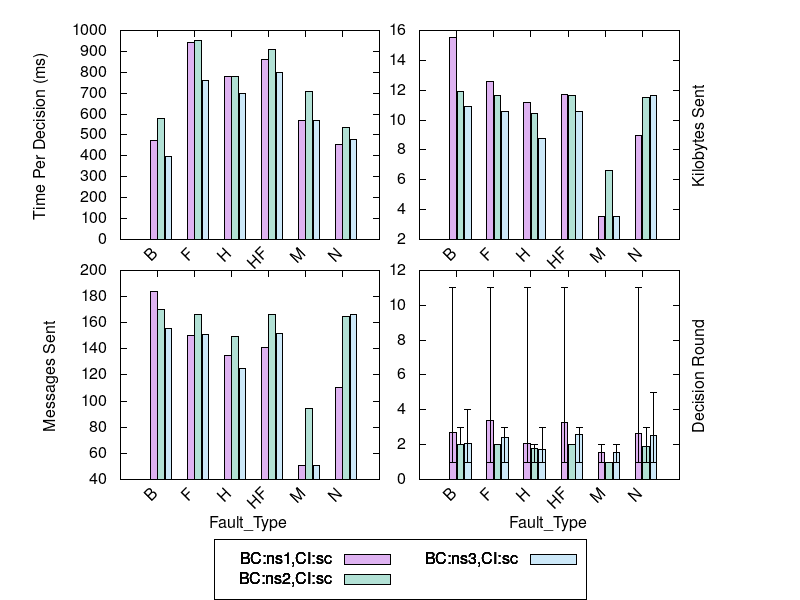}
  \end{center}
  \caption{16 node geo-distributed experiment of the non-signature based algorithms using coin presets where 1/3 of the nodes exhibit various
    faulty behaviors and approximately 1/2 of the non-faulty nodes propose 1.}
  \label{fig:ns-byz-50}
\end{figure*}

\begin{figure*}[ht!]
\begin{center}
    \includegraphics[scale=0.50]{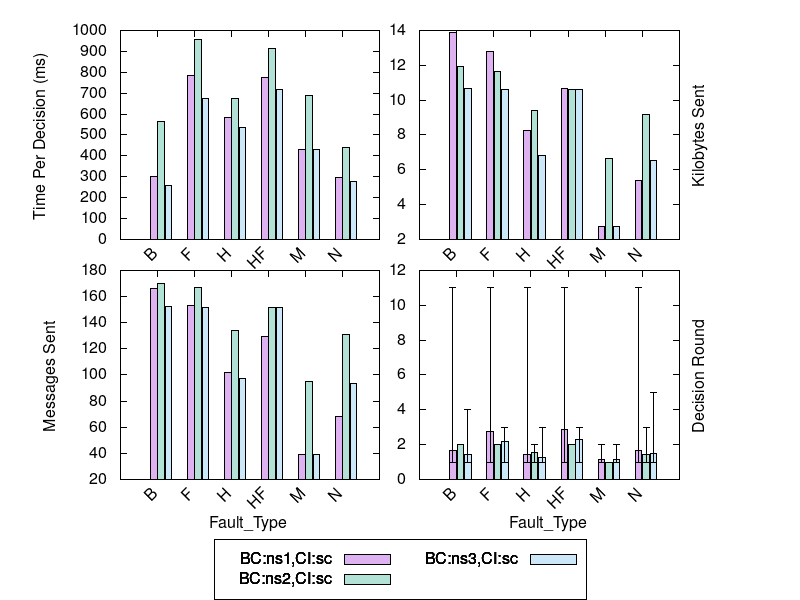}
  \end{center}
  \caption{16 node geo-distributed experiment of the non-signature based algorithms using coin presets where 1/3 of the nodes exhibit various
    faulty behavior and approximately 2/3 of the non-faulty nodes
  propose 1.}
  \label{fig:ns-byz-66}
\end{figure*}

Figures~\ref{fig:s-byz-33}-\ref{fig:ns-byz-66} present the results of the experiments using
the same $16$ node geo-distributed configuration with the faulty behaviors.
Faulty nodes are distributed evenly across different regions.
Measurements are taken from the non-faulty nodes.

Figure~\ref{fig:s-byz-33} presents the results of the experiment using the signature based algorithms with coin presets where
approximately $1/3$ of non-faulty nodes propose $1$.
For nearly all fault types the decision time is increased compared to when using the non-faulty (N) behavior.
The flip behavior has the highest time to decide at nearly $700$ milliseconds for BC:s2 compared to just over $300$
milliseconds for the normal behavior.

For the flip (F), half-half (H), and half-half fixed (HF) behaviors the number of rounds to decide is higher that in the normal case, accounting
for the additional decision time and message and bit complexity.
The both (B) behavior seems to have minimal affect on BC:s2 and BC:s3 as in this case the nodes will receive close to $n$ messages
with value $0$ allowing them to decide quickly.

Even though the mute (M) behavior reduces the message count, the bytes sent, and has low values for the decision round,
the time to decide is still higher than the non-faulty behavior. This is likely due to non-faulty nodes having to wait
for messages from non-faulty nodes that have high network latency.

The variation in performance with different faulty behaviors for BC:s1 is smaller than for the other algorithms,
which is reflected by the fact that BC:s1 has the most stable round to decide value throughout the experiments.
This may be caused by the shorter rounds of BC:s1, thus providing fewer possibilities for variation, i.e.
in a single round only the value of the coin can be decided.

Figure~\ref{fig:s-byz-50} presents the results of the experiment using the signature based algorithms with coin presets where
approximately $1/2$ of non-faulty nodes propose $1$.
Like in the previous experiment, the highest time to decide is with BC:s2 at around $700$ milliseconds with similar
round and messaging costs.

With the both (B) behavior, BC:s3 is able to terminate even faster than with the non-faulty (N) behavior.
This is due to enough $1$ values being broadcast that it is able to terminate quickly in round $1$.
A similar effect seems to be happening with the half-half (H) behavior for both BC:s1 and BC:s3.

Figure~\ref{fig:s-byz-66} presents the results of the experiment using the signature based algorithms with coin presets where
approximately $2/3$ of non-faulty nodes propose $1$.
While generally these experiments show similar patterns to the previous experiments, the main difference is
that there are enough $1$ values being proposed so that BC:s1 and BC:s3 are able to decide in round $1$
in many cases, allowing them to have better performance.

Figure~\ref{fig:ns-byz-33} presents the results of the experiment using the non-signature based algorithms with coin presets where
approximately $1/3$ of non-faulty nodes propose $1$.
Like the signature algorithms, all faulty behaviors increase the decision time of the algorithms.
This can be principally attributed to the fact that the faulty behaviors are able to increase the decision round by increasing
the amount of disagreement in the system.

The worst case decision times are higher than the signature algorithms given that the non-signature based algorithms perform
more message steps per round.
The highest decision time comes from the flip (F) fault type with BC:s1 having decision time at over $1000$ milliseconds.
Furthermore, the non-signature based algorithms have higher message count than the signature based algorithms, but
still send less bytes due not having the signature overhead.
In the both (B) fault type, the message count is the highest as in this case both $0$ and $1$ become valid
most frequently, and the nodes end up performing the echo message pattern.

Figure~\ref{fig:ns-byz-50} presents the results of the experiment using the non-signature based algorithms with coin presets where
approximately $1/2$ of non-faulty nodes propose $1$.
Here the main difference from the previous experiment is that algorithms BC:s1 and BC:s3 have improved performance due
to the higher percentage of $1$'s being proposed allowing the algorithms to more frequently decide in the first round.

Figure~\ref{fig:ns-byz-66} presents the results of the experiment using the non-signature based algorithms with coin presets where
approximately $2/3$ of non-faulty nodes propose $1$.
Again algorithms BC:s1 and BC:s3 have improved performance due
to the higher percentage of $1$'s being proposed allowing them to more frequently decide in the first round.
Note that the fastest decision times of BC:ns1 and BC:ns3 match the fastest decision times of BC:s1 and BC:s3
given that in both cases the fastest decision can happen after $2$ message broadcasts.

\subsection{Increased number of nodes}
\begin{figure*}[ht!]
\begin{center}
    \includegraphics[scale=0.50]{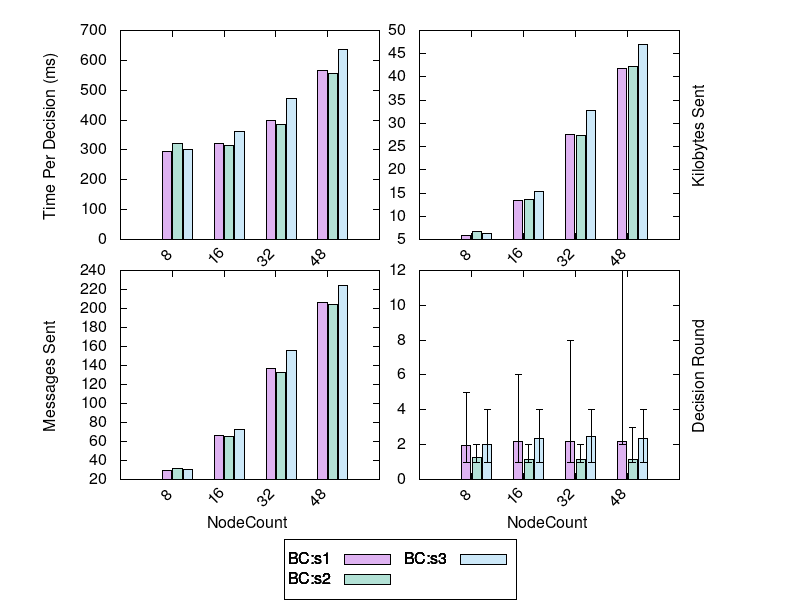}
  \end{center}
  \caption{Geo-distributed experiment of the signature based algorithms with coin presets where approximately 1/3 of the nodes
  propose 1 using different number of nodes.}
  \label{fig:s-sca-33}
\end{figure*}

\begin{figure*}[ht!]
\begin{center}
    \includegraphics[scale=0.50]{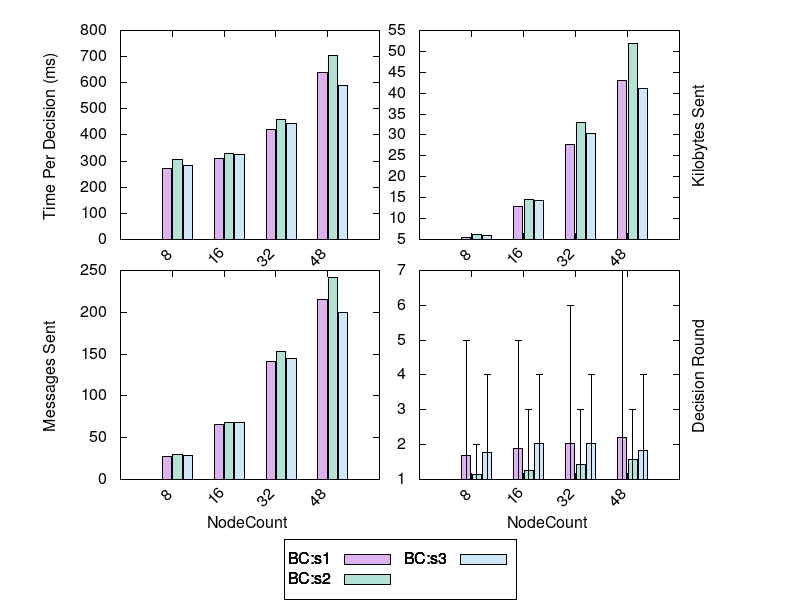}
  \end{center}
  \caption{Geo-distributed experiment of the signature based algorithms where approximately 1/2 of the nodes
  propose 1 using different number of nodes.}
  \label{fig:s-sca-50}
\end{figure*}

\begin{figure*}[ht!]
\begin{center}
    \includegraphics[scale=0.50]{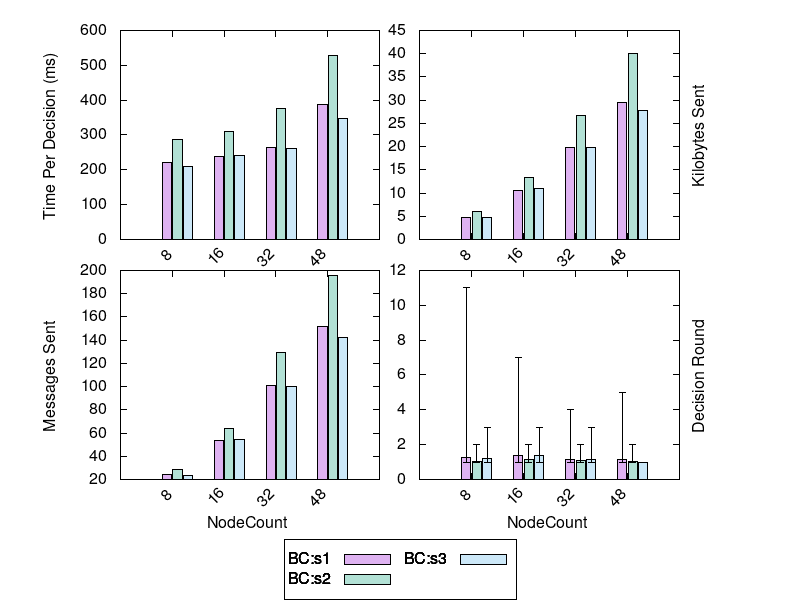}
  \end{center}
  \caption{Geo-distributed experiment of the signature based algorithms where approximately 2/3 of the nodes
  propose 1 using different number of nodes.}
  \label{fig:s-sca-66}
\end{figure*}

\begin{figure*}[ht!]
\begin{center}
    \includegraphics[scale=0.50]{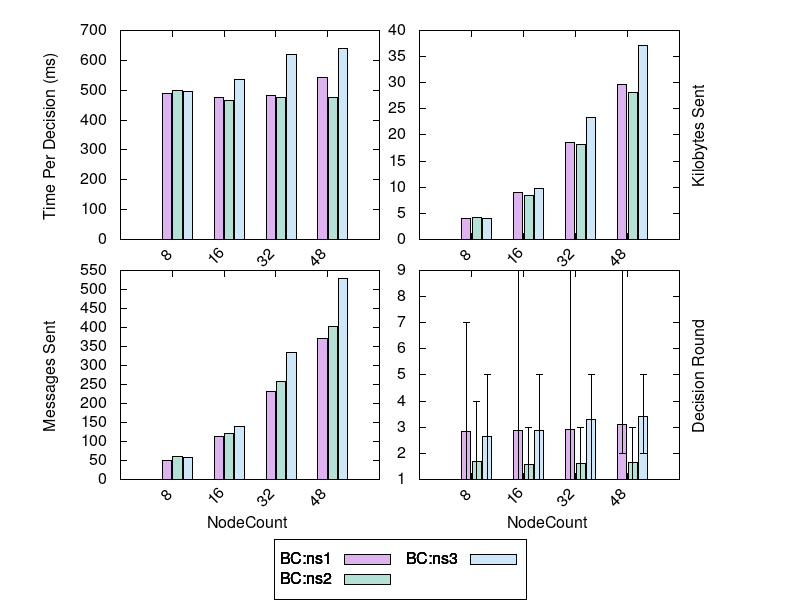}
  \end{center}
  \caption{Geo-distributed experiment of the non-signature based algorithms where approximately 1/3 of the nodes
  propose 1 using different number of nodes.}
  \label{fig:ns-sca-33}
\end{figure*}

\begin{figure*}[ht!]
\begin{center}
    \includegraphics[scale=0.50]{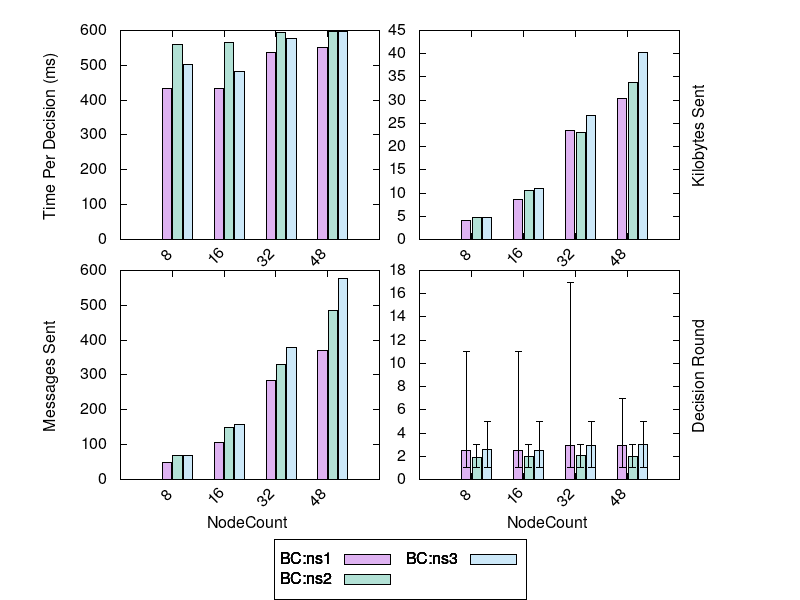}
  \end{center}
  \caption{Geo-distributed experiment of the non-signature based algorithms where approximately 1/2 of the nodes
  propose 1 using different number of nodes.}
  \label{fig:ns-sca-50}
\end{figure*}

\begin{figure*}[ht!]
\begin{center}
    \includegraphics[scale=0.50]{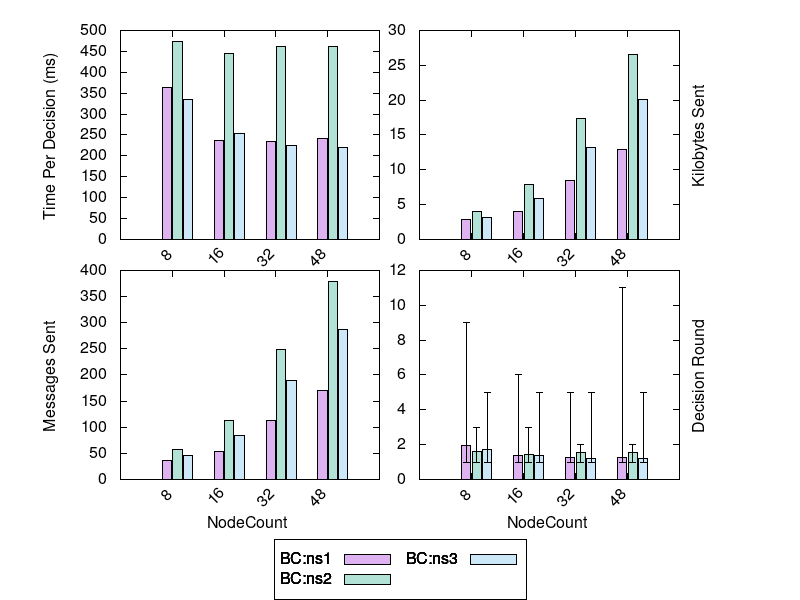}
  \end{center}
  \caption{Geo-distributed experiment of the non-signature based algorithms where approximately 2/3 of the nodes
  propose 1 using different number of nodes.}
  \label{fig:ns-sca-66}
\end{figure*}

Figures~\ref{fig:s-sca-33}-\ref{fig:ns-sca-66} present the results of experiments using
$8$, $16$, $32$, and $48$ n1-standard-2 Google cloud compute instances distributed across the same eight regions used in
the previous experiments.

Figure~\ref{fig:s-sca-33} presents the results of the experiment using the signature based algorithms with coin presets where
approximately $1/3$ of non-faulty nodes propose $1$.
All algorithms exhibit fairly similar performance with latencies increasing from $300$ millisecond at $8$ nodes
to around $600$ milliseconds at $48$ nodes.
Algorithm BC:s3 has  higher overhead at higher number of nodes due to a small increase in decision round.
The number of bytes and messages sent per node grows linearly as expected, but at $45$ kilobytes sent per consensus
instance the load is still quite low.
Thus, the increase in latency is largely coming form the increased cryptographic computation overhead.

Figure~\ref{fig:s-sca-50} presents the results of the experiment using the signature based algorithms with coin presets where
approximately $1/2$ of non-faulty nodes propose $1$.
Algorithms BC:s1 and BC:s3 show a small increase in performance due to them being more frequently able to decide
in the first round due to the increase in $1$ proposals.

Figure~\ref{fig:s-sca-66} presents the results of the experiment using the signature based algorithms with coin presets where
approximately $2/3$ of non-faulty nodes propose $1$.
Again the performance of BC:s1 and BC:s3 benefit most from the increase of $1$ proposals.

Figure~\ref{fig:ns-sca-33} presents the results of the experiment using the non-signature based algorithms with coin presets where
approximately $1/3$ of non-faulty nodes propose $1$.
Here the decision time of BC:ns3 increases from $500$ milliseconds at $8$ nodes to $650$ milliseconds at $48$ nodes.
This appears to be caused by a small increase in the average decision round.
The performance of BC:ns1 shows a similar, but smaller effect, while the decision time of BC:ns2 stays around $500$ milliseconds.
While the increase in decision round could be caused by a higher level of asynchrony in the system due to the increased number
of nodes, it is not clear why BC:ns3 is most affect by this.

Figure~\ref{fig:ns-sca-50} presents the results of the experiment using the non-signature based algorithms with coin presets where
approximately $1/2$ of non-faulty nodes propose $1$.
In this case the decision time of BC:ns2 is about $100$ milliseconds
slower than when $1/3$ of the nodes propose $1$, where the increase in disagreement
in proposals is causing a higher decision round.
Differently, algorithms BC:ns1 and BC:ns2 have approximately $100$ millisecond decrease in decision time due to
them being able to more frequently decide in the first round due to the increase in $1$ proposals.

Figure~\ref{fig:ns-sca-66} presents the results of the experiment using the non-signature based algorithms with coin presets where
approximately $2/3$ of non-faulty nodes propose $1$.
Interestingly, the decision time of BC:ns1 and BC:ns3 decreases with an increase in the number of nodes,
with BC:ns3 reaching as low as $220$ milliseconds with $48$ nodes.
While this is caused by a decrease in average decision round as the number of nodes increase, it is
not clear why the round decrease happens and may be an artifact of the benchmark configuration.
The time to decide of BC:ns3 stays stable at around $450$ milliseconds.

\section{Discussion}
Each algorithm has cases where it performs better than all the other algorithms for certain measurements
so choosing an algorithm will depend on the experimental environment.
The use of signatures provide advantages that may be applicable beyond just the consensus algorithm
such as using irrefutably to discourage Byzantine behavior, while the non-signature based algorithms
provide many performance benefits.
Some points to consider may be:
\begin{itemize}
\item The non-signature algorithms scale much better as they use much less computational resources
  and less network bandwidth.
\item When using a low latency network the performance advantage of the non-signature based algorithms
  can be quite significant given the computational overhead of signatures.
\item Signature based algorithms have a ``sweet spot'' where due to having a lower message step complexity (in most cases)
  than the non-signature based algorithms they will decide faster in networks that have high enough latency
  to compensate for the time needed for the cryptographic operations for the given number of nodes.
  Otherwise the non-signature based algorithms will decide faster.
\item The minimum number of messages to decide is $2$ for both the signature and non-signature based algorithms.
  This will often be the expected message pattern when using multi-valued reductions.
\end{itemize}

The algorithms BC:s1 and BC:ns1 have the highest variance in performance due to the fact that
they can only decide the value of the coin in a round and perform best when using coin presets.
Algorithms BC:s2 and BC:ns2 provide fairly stable performance.
Algorithms BC:s3 and BC:ns3 provide the benefits of both with a small overhead in some cases.

While simple Byzantine behaviors do have an effect on performance,
the experiments still show that even in networks with high latency variance (from $2$ to $280$ milliseconds), where nodes
have high initial disagreement, 
the algorithms are still able to decide in the first round without computing the value of a coin.
This seems to suggest that a powerful network adversary is needed reduce performance further, but as the expected
termination round is still a small constant value the performance should still be comparable.

\paragraph{Future work.}
It should be interesting to investigate the performance of the algorithms when using adversarial
networks, but while still using realistic models.
Furthermore the algorithms should be tested when used as part of various multi-value reductions.


\begin{thebibliography}{99}
\footnotesize{  

\bibitem{A03}
James Aspnes. Randomized protocols for asynchronous consensus.
{\it Distrib. Comput.}, 16(2-3):165-175, September 2003.


\bibitem{BO83}
Michael Ben-Or. Another advantage of free choice (extended abstract):
Completely asynchronous agreement protocols. {\it In Proceedings of the
Second Annual ACM Symposium on Principles of Distributed Computing}, PODC '83, pages 27-30, 1983

\bibitem{BG93}
Berman P. and Garay J.A., Randomized distributed agreement revisited.
{\it 33rd Annual Int'l Symposium on Fault-Tolerant Computing (FTCS' 93)},
IEEE Computer Press, pp. 412-419, 1993.

\bibitem{BSA14}
Alyson Bessani, Joao Sousa, and Eduardo E. P. Alchieri. State machine
replication for the masses with bft-smart. {\it In 2014 44th Annual IEEE/IFIP
International Conference on Dependable Systems and Networks}, pages
355-362, June 2014.

\bibitem{BLS04}
Dan Boneh, Ben Lynn, and Hovav Shacham. 2004. Short Signatures from the Weil Pairing.
{\it J. Cryptol. 17, 4} (September 2004), 297-319.

\bibitem{blake}
BLAKE2 Hashing. \url{https://blake2.net/}

\bibitem{B83}
Bracha G.,Asynchronous Byzantine agreement protocols. {\it Information \& Computation}, 1987.

\bibitem{B84}
Bracha G., An asynchronous $(n-1)/3$-resilient consensus protocol. {\it Proc. 3rd Annual ACM Symposium on Principles of Distributed Computing (PODC'84)}, 1984.

\bibitem{B87}
Gabriel Bracha. An o(log n) expected rounds randomized byzantine
generals protocol. {\it J. ACM}, 34(4):910-920, October 1987

\bibitem{BT83}
Gabriel Bracha and Sam Toueg. Asynchronous consensus and byzantine protocols in faulty environments. Technical Report TR83-559,
Cornell University, 1983.




\bibitem{CKS05}
Christian Cachin, Klaus Kursawe, and Victor Shoup. Random oracles
in constantinople: Practical asynchronous byzantine agreement using
cryptography. {\it Journal of Cryptology}, 18(3):219-246, 2005.

\bibitem{CGR11}
Cachin Ch., Guerraoui R., and Rodrigues L., {\it Reliable and secure distributed programming, Springer}, 2011.

\bibitem{CR93}
Ran Canetti and Tal Rabin. Fast asynchronous byzantine agreement
with optimal resilience. In Proceedings of the Twenty-fifth Annual ACM
Symposium on Theory of Computing, {\it STOC '93}, pages 42-51, 1993.

\bibitem{CL02}
 Miguel Castro and Barbara Liskov. Practical byzantine fault tolerance
and proactive recovery. {\it ACM Trans. Comput}. Syst., 20(4):398-461,
November 2002.

\bibitem{CKS20}
Cohen, Shir, Idit Keidar, and Alexander Spiegelman. Not a COINcidence: Sub-Quadratic Asynchronous Byzantine Agreement WHP. {\it arXiv preprint arXiv:2002.06545}, 2020.

\bibitem{C20}
Tyler Crain. A Simple and Efficient Asynchronous Randomized Binary Byzantine Consensus Algorithm.
{\it arXiv preprint arXiv:2002.04393}, 2020.
  
\bibitem{C220}
Tyler Crain. A Simple and Efficient Binary Byzantine Consensus Algorithm
using Cryptography and Partial Synchrony. {\it arXiv preprint arXiv:2001.07867}, 2020.

\bibitem{C320}
  Tyler Crain. Two More Algorithms for Randomized Signature-Free Asynchronous Binary Byzantine Consensus with $t < n/3$ and $O(n^2)$ Messages and $O(1)$ Round Expected Termination.  {\it arXiv preprint arXiv:2002.08765}, 2020.

\bibitem{CGLR18}
Tyler Crain, Vincent Gramoli, Mikel Larrea, and Michel Raynal.
Dbft: Efficient leaderless byzantine consensus and its applications to
blockchains. {\it In Proceedings of the 17th IEEE International Symposium
on Network Computing and Applications (NCA'18)}. IEEE, 2018.




\bibitem{DDS87}
Danny Dolev, Cynthia Dwork, and Larry Stockmeyer. On the minimal
synchronism needed for distributed consensus. {\it J. ACM}, 34(1):77-97,
January 1987.


\bibitem{DRZ18}
Sisi Duan, Michael K. Reiter, and Haibin Zhang. BEAT: Asynchronous BFT Made Practical. {\it In Proc. of the Conference on Computer and Communications Security (CCS'18)}. 2018.


\bibitem{DLS88}
Cynthia Dwork, Nancy A. Lynch, and Larry J. Stockmeyer. Consensus
in the presence of partial synchrony. {\it J. ACM}, 35(2):288-323, 1988.

\bibitem{FM97}
PESECH FELDMAN and SILVIO Micali. An optimal probabilistic
protocol for synchronous byzantine agreement. {\it SIAM J. Computing},
26(4):873-933, 1997.

\bibitem{FMR05} Friedman R., Most\'efaoui A., and Raynal M., Simple and efficient oracle-based consensus protocols for asynchronous Byzantine systems. {\it IEEE Transactions on Dependable and Secure Computing}, 2005.


\bibitem{FS86}
A. Fiat and A. Shamir.  How to prove yourself:  Practical solutions to identification andsignature  problems.
{\em Advances  in  Cryptology: CRYPTO  '86}, volume 263. Springer, 1987.

\bibitem{FL82}
Fischer M.J. and Lynch N.A.,
A lower bound for the time to assure interactive consistency.
{\em Information Processing Letters}, 14(4):183-186 (1982)


\bibitem{FLP85}
Fischer M.J., Lynch N.A.,  and Paterson M.S.,
Impossibility of distributed consensus with one faulty process.
{\em Journal of the ACM}, 32(2):374-382 (1985)

\bibitem{FMR07}
Friedman R., Most\'efaoui A., Rajsbaum S., and Raynal M., Distributed agreement problems and their
connection with error-correcting codes. {\it IEEE Transactions on Computers}, 56(7):865-875, 2007. 

\bibitem{FP90}
Oded Goldreich and Erez Petrank. The best of both worlds: Guaranteeing termination in fast randomized byzantine agreement protocols.
{\it Inf. Process. Lett.}, 36(1):45-49, 1990.

\bibitem{GO}
  Go Programming Language. \url{https://golang.org/}.
  

\bibitem{nacl}
  Go NaCl Implementation. \url{https://godoc.org/golang.org/x/crypto/nacl}.
  
\bibitem{KS16}
  Valerie King and Jared Saia. Byzantine agreement in expected polynomial time. {\it J. ACM}, 63(2):13, 2016.

\bibitem{kyber}
  Kyber library, Advanced crypto library for the Go language. \url{https://github.com/dedis/kyber}.

\bibitem{LSP82}
Leslie Lamport, Robert Shostak, and Marshall Pease. The byzantine
generals problem. {\it ACM Trans. Program. Lang. Syst.}, 4(3):382-401, July
1982.

\bibitem{LVCQV16}
Shengyun Liu, Paolo Viotti, Christian Cachin, Vivien Qu\'ema, and
Marko Vukolic. XFT: practical fault tolerance beyond crashes. {\it In 12th
USENIX Symposium on Operating Systems Design and Implementation,
OSDI 2016}, Savannah, GA, USA, November 2-4, 2016., pages 485-500,
2016.

\bibitem{M18}
Ethan MacBrough. Cobalt: BFT Governance in Open Networks.
{\it arXiv preprint arXiv:1802.07240}, 2018.

\bibitem{MA06}
Jean-Philippe Martin and Lorenzo Alvisi. Fast byzantine consensus.
{\it IEEE Trans. Dependable Sec. Comput.}, 3(3):202-215, 2006.

\bibitem{MRV99}
Silvio Micali, Michael Rabin, and Salil Vadhan. Verifiable random functions. {\it In Foundations of Computer Science}, 1999.

\bibitem{MXC16}
Miller, A., Xia, Y., Croman, K., Shi, E., Song, D. The honey badger of BFT protocols. {\it In Proceedings of the Conference on Computer and Communications Security}. 2016.

\bibitem{MNCV06}
 Moniz, H., Neves, N. F., Correia, M., Verissimo, P. Experimental comparison of local and shared coin randomized consensus protocols. {\it In 2006 25th IEEE Symposium on Reliable Distributed Systems}. 2006.


\bibitem{MMR14}
Achour Most\'efaoui, Hamouma Moumen, and Michel Raynal.
Signature-free asynchronous byzantine consensus with $T < N /3$ and
$O(N^2)$ messages. {\it In Proceedings of the 2014 ACM Symposium on Principles of Distributed Computing, PODC '14},
pages 2-9, New York, NY,
USA, 2014. ACM.

\bibitem{MMR15}
Achour Mostéfaoui, Hamouma Moumen, and Michel Raynal.
Signature-Free Asynchronous Binary Byzantine Consensus with t < n/3, O(n2) Messages,
and O(1) Expected Time. {\it J. ACM 62, 4}. Article 31. 2015.


\bibitem{MR17}
Achour Most\'efaoui and Michel Raynal. Signature-free asynchronous byzantine systems: from multivalued to binary consensus with
$t < n/3$, $O(n^2)$ messages, and constant time. {\it Acta Informatica, 2017}.
Accepted: 19 April 2016

\bibitem{MRT00}
Achour Most\'efaoui, Michel Raynal, and Fr\'ed\'eric Tronel. From binary
consensus to multivalued consensus in asynchronous message-passing
systems. {\it Inf. Process. Lett., 73(5-6):207-212}, March 2000.


\bibitem{NCV05}
N. F. Neves, M. Correia, and P. Verissimo. Solving vector consensus
with a wormhole. {\it IEEE Trans. on Parallel and Distributed Systems},
16(2):1120-1131, 2005.

\bibitem{PCR14}
Arpita Patra, Ashish Choudhury, and C. Pandu Rangan. Asynchronous
byzantine agreement with optimal resilience. {\it Distributed Computing},
27(2):111-146, 2014.

\bibitem{PSL80}
M. Pease, R. Shostak, and L. Lamport. Reaching agreement in the
presence of faults. {\it J. ACM}, 27(2):228-234, April 1980

\bibitem{R83}
Michael O. Rabin. Randomized byzantine generals. {\it In Proceedings of
the 24th Annual Symposium on Foundations of Computer Science, SFCS
'83}, pages 403-409, 1983.

\bibitem{S79}
A. Shamir. How to share a secret. {\it Communications of the ACM}, 1979.


\bibitem{ST87}
Srikanth T.K. and Toueg S., Simulating authenticated broadcasts to derive simple fault-tolerant algorithms.
{\it Distributed Computing}, 2:80-94, 1987.


\bibitem{T84}
Sam Toueg. Randomized byzantine agreements. {\it In Proceedings of the
Third Annual ACM Symposium on Principles of Distributed Computing,
PODC '84}, pages 163-178, 1984.

\bibitem{TC84}
  Russell Turpin and Brian A. Coan. Extending binary byzantine agreement to multivalued byzantine agreement.
  {\it Inf. Process. Lett.}, 18(2):73-
76, 1984.

\bibitem{ZC09}
Jialin Zhang and Wei Chen. Bounded cost algorithms for multivalued
consensus using binary consensus instances. {\it Information Processing
Letters}, 109(17):1005-1009, 2009.


}
\end{thebibliography}
\end{document}